\documentclass[fleqn,usenatbib]{mnras}

\usepackage[T1]{fontenc}
\usepackage{physics}
\usepackage{tabularx}
\DeclareRobustCommand{\VAN}[3]{#2}
\let\VANthebibliography\thebibliography
\def\thebibliography{\DeclareRobustCommand{\VAN}[3]{##3}\VANthebibliography}
\defcitealias{Kathirgamaraju_2019}{KGB19}

\usepackage{graphicx}	
\usepackage{amsmath}	
\usepackage{amssymb}	
\usepackage{subcaption}
\usepackage{newtxtext,newtxmath}

\title[Non-thermal emission of BNS merger]{Non-thermal emission from mildly relativistic dynamical ejecta of neutron star mergers}

\author[G. Sadeh, et al.]{
Gilad Sadeh\thanks{E-mail: gilad.sade@weizmann.ac.il}, Or Guttman,
Eli Waxman
\\
\it{Dept. of Particle Phys. \& Astrophys., Weizmann Institute of Science, Rehovot 76100, Israel}
}

\date{Accepted XXX. Received YYY; in original form ZZZ}

\pubyear{2022}

\begin{document}
\label{firstpage}
\pagerange{\pageref{firstpage}--\pageref{lastpage}}
\maketitle

\begin{abstract}
Binary neutron star mergers are expected to produce fast dynamical ejecta, with mildly relativistic velocities extending to $\beta=v/c>0.6$. We consider the radio to X-ray synchrotron emission produced by collisionless shocks driven by such fast ejecta into the interstellar medium. Analytic expressions are given for spherical ejecta with broken power-law mass (or energy) distributions, $M(>\gamma\beta)\propto(\gamma\beta)^{-s}$ with $s=s_{\rm KN}$ at $\gamma\beta<\gamma_0\beta_0$ and $s=s_{\rm ft}$ at $\gamma\beta>\gamma_0\beta_0$ (where $\gamma$ is the Lorentz factor). For parameter values characteristic of merger calculation results- a "shallow" mass distribution, $1<s_{\rm KN}<3$, for the bulk of the ejecta (at $\gamma\beta\approx 0.2$), and a steep, $s_{\rm ft}>5$, "fast tail" mass distribution- our model provides an accurate (to 10's of percent) description of the evolution of the flux, including at the phase of deceleration to sub-relativistic expansion. This is a significant improvement over earlier results, based on extrapolations of results valid for $\gamma\beta\gg1$ or $\ll1$ to $\gamma\beta\approx1$, which overestimate the flux by an order of magnitude for typical parameter values. It will enable a more reliable inference of ejecta parameters from future measurements of the non-thermal emission. For the merger event GW170817, the existence of a "fast tail" is expected to produce detectable radio and X-ray fluxes over a time scale of $\sim10^4$days.
\end{abstract}




\section{Introduction}

Mergers of binary neutron stars (BNS) are expected to produce ejecta expanding at mildly relativistic velocities, $\gamma\beta>0.1$, through several different mechanisms including dynamic ejection by tidal interaction and shocks, neutrino driven winds and (secular, spiral wave) mass ejection from a disk formed during the merger \citep[see][for reviews]{Fernandez_2016,radice_2020}. The amount of ejected mass and its velocity distribution depend on the parameters of the merging system and on the equation of state (EoS) of the nuclear matter. Measurements of the electromagnetic emission generated by the expanding ejecta will enable us to constrain its properties, and hence the parameters of the merging binary and of the EoS.

Observations of the optical (UV-IR) electromagnetic counterpart of GW170817 are consistent with an ejecta of $\approx 0.05M_\odot$ expanding at $\beta>0.1$ and undergoing continuous heating by radioactive decays \citep{Arcavi_2017,Nicholl_2017,Metzger_2017}. This is remarkably consistent with the "kilonova" emission predicted to follow NS mergers \citep{LiPac98}. However, the detailed properties of the ejecta, that are inferred from observations, are inconsistent with those obtained in merger simulations \citep[see e.g.][for a recent detailed discussion]{nedora_2021b}. In particular, the mass of the ejecta is larger than obtained in simulations, and it is difficult to explain the existence of a fairly massive, $\simeq0.05M_\odot$, fast, $v\sim0.3c$, component with low opacity corresponding to a large initial electron fraction $Y_e$, that is suggested by the early UV/blue emission \citep{Waxman_2018}.

Numeric merger calculations find that the velocity distribution of the dynamical ejecta may extend to large velocities, well above the characteristic velocity, $\beta\sim0.1-0.2$, of the bulk of the ejecta that is responsible for the generation of the UV-IR "kilonova" (KN) emission \citep[e.g.][]{Bauswein_2013,Ishii_2018,Radice_2018,shibata_2019, ciolfi_2020, bernuzzi_2020, nedora_2021a,Dean_2021, Hajela_2022}. The existence of such a fast ejecta "tail" has recently been suggested as an explanation of a possible "re-brightening" of the X-ray emission of the GW170817 electromagnetic counterpart \citep{nedora_2021a,Hajela_2022}. However, the evidence for rebrightening is not conclusive \citep{Balasubramanian_2021,Balasubramanian_2022,Troja_2022,Oconnor_2022}. The mass and velocity profile of the "fast tail" depend strongly on the parameters of the binary system and on the EoS. Depending on parameter values, the ejecta mass at $\beta>0.6$ varies, for example, between $10^{-7}M_\odot$ and $10^{-4}M_\odot$, and the maximal velocity varies from $\gamma\beta\approx0.6$ to $\gamma\beta>3$. This implies that measurements of the non-thermal emission driven by the fast ejecta may provide stringent constraints on model parameters.

Two points should, however, be noted here. First, a reliable determination of constraints would require a significant reduction of the current large numerical uncertainty in the numerically calculated fast tail parameters, as reflected by the large variations of results between different simulations \citep[e.g.][]{Dean_2021}.
Second, if BNS mergers are the sources of (short) gamma-ray bursts \citep[see][for reviews]{2002MeszGRBrev,2004PiranGRBrev,2007NakarSGRBrev}, then they are expected to launch highly relativistic jets. Depending on the jet axis with respect to our line of sight, and on the ratio of energies carried by the jet and by the dynamical ejecta, the emission of non-thermal radiation driven by the jet may dominate that driven by the dynamical ejecta. In the case of GW170817, the rapid decline of the observed non-thermal emission \citep{Troja_2019,fong_2019,lamb_2019,hajela_2019,Troja_2020,Makhathini_2020,Nakar_2020} and the super-luminal motion of the radio centroid suggest that the observed non-thermal emission is driven by a relativistic jet \citep[the presence of plasma expanding at $\gamma>10$ cannot be directly inferred from observations, which may be accounted for by a $\gamma\sim4$ jet][]{Mooley_2018}. In general, due to the lower mass and higher Lorenz factor of the jet, we may expect the emission to be dominated at late time by the dynamical ejecta. The interpretation of future observations will require disentangling the two components.

We consider in this paper the radio to X-ray synchrotron emission produced by collisionless shocks driven by mildly relativistic ejecta expanding into an interstellar medium (ISM) with a uniform (number) density $n$. Analytic expressions are derived for the case of spherical ejecta (the effects of deviations from spherical symmetry will be discussed in a follow-up paper) with a broken power-law dependence of mass on momentum,
\begin{equation}
\label{eq:profile}
    M(>\gamma\beta)= M_0
    \begin{cases}
    \left(\frac{\gamma\beta}{\gamma_0\beta_0}\right)^{-s_\text{ft}} & \text{for}\quad \gamma_0\beta_0<\gamma\beta,\\
    \left(\frac{\gamma\beta}{\gamma_0\beta_0}\right)^{-s_\text{KN}} & \text{for}\quad 0.1<\gamma\beta<\gamma_0\beta_0,
    \end{cases}
\end{equation}
with parameter values characteristic of the results of numerical calculations of the dynamical ejecta \citep[e.g.][]{Radice_2018,Hajela_2022,nedora_2021a,Fujibayashi_2022,Rosswog_2022}- a "shallow" mass distribution, $1<s_{\rm KN}<3$, for the bulk of the mass (at $\gamma\beta\approx 0.2$) and a steep, $s_{\rm ft}>5$, fast tail mass distribution \citep[see, e.g. figure 10 of][for a useful description of mass profiles obtained in various calculations]{Hajela_2022}. We note that the "shallow" dependence of mass on velocity at $\gamma\beta<0.3$ is consistent with the early evolution of the photospheric radius of the emission following GW170817, which implies $M(>\gamma\beta)\propto(\gamma\beta)^{-s_{\rm KN}}$ with $s_{\rm KN}\approx1.6$ at $0.1<\gamma\beta<0.3$ \citep{Waxman_2018}. Analytic expressions for a broken power-law dependence of ejecta energy on momentum,
\begin{equation}
\label{eq:profileE}
    E(>\gamma\beta)= E_0
    \begin{cases}
    \left(\frac{\gamma\beta}{\gamma_0\beta_0}\right)^{-\alpha_\text{ft}} & \text{for}\quad \gamma_0\beta_0<\gamma\beta,\\
    \left(\frac{\gamma\beta}{\gamma_0\beta_0}\right)^{-\alpha_\text{KN}} & \text{for}\quad 0.1<\gamma\beta<\gamma_0\beta_0,
    \end{cases}
\end{equation}
are also provided ($\alpha_\text{ft}>3,0<\alpha_\text{KN}<1.5$), based on the results obtained for a broken power-law dependence of mass on momentum.

The non-thermal flux is derived assuming that fractions $\varepsilon_e$ and $\varepsilon_B$ of the post-shock internal energy density are carried by non-thermal electrons and magnetic fields respectively, and the electrons are assumed to follow a power-law energy distribution, $dn_e/d\gamma_e\propto \gamma_e^{-p}$, where $\gamma_e$ is the electron Lorenz factor (at the plasma rest-frame) and $2\leq p\leq2.5$. This phenomenological description of the post-shock plasma conditions is supported by a wide range of observations and plasma calculations, for both relativistic and non-relativistic shocks \citep[see][for reviews]{Blandford_1987,Waxman_2006,Bykov_2011,Sironi_2013}.

The accuracy of the analytic approximations derived in this paper for the non-thermal flux is verified through comparison to numerical calculations for a wide range of parameter values. It is significantly improved compared to that of earlier work, mainly due to an improved treatment of the dynamics of the ejecta at the relevant velocity range, $0.1<\gamma\beta<3$. The accuracy of the approximate description of the dynamics is important due to the strong dependence of the flux on the velocity of the emitting plasma (approx. $\propto(\gamma\beta)^4$, see below). In addition, we provide an approximate analytic description for both the phase of rising flux, which was addressed in earlier papers \citep{Nakar_2011,Kathirgamaraju_2019}, and for the phase of flux decline, which was not addressed in detail before.

The paper is organized as follows. The analytic expressions providing an approximate description of the non-thermal flux are derived in \S~\ref{sec:model}. Their accuracy is estimated by a comparison to the results of numeric relativistic-hydrodynamic calculations in \S~\ref{sec:numeric}. Our results are compared to those of earlier work in \S~\ref{sec:earlier}. The implications of our results for future observations of the electromagnetic counterpart of GW170817 are discussed in \S~\ref{sec:GW170817}, and our conclusions are summarized in \S~\ref{sec:summary}.

\section{Analytic model}
\label{sec:model}

As the ejecta expands, a forward shock is driven into the ISM and a reverse shock is driven into the ejecta (see Fig.~\ref{fig:FRshocks}). The kinetic energy of the ejecta decreases due to its deceleration by the reverse shock, and the energy is transferred to the shocked ISM plasma. The shocked ISM mass increases with the expanding shock radius, while its energy density decreases due to shock deceleration. The increase of the mass of hot plasma acts to increase the non-thermal emission, while the deceleration acts to suppress it (due both to the decreasing energy density and to the decreasing effect of flux enhancement due to relativistic expansion). For sufficiently steep drop of mass with velocity, $s_\gtrsim5$ as shown below, the former effect dominates and the non-thermal flux increases with time. This is due to the fact that the deceleration is slower for a steeper mass distribution, since the shocked ISM mass that needs to be accumulated in order to decelerate the shock below a certain velocity increases faster with decreasing velocity (as the ejecta mass increases faster with decreasing velocity).

For the steep mass profiles of the fast tail, inferred from numeric calculations, we thus expect the non-thermal flux to increase with time as long as the reverse shock propagates through the fast tail of the ejecta. Once the reverse shock penetrates into the bulk of the ejecta, where the mass distribution as function of velocity is shallow, the deceleration becomes more rapid and the flux begins to decrease. At late time, after the reverse shock crosses the entire ejecta and the expansion velocity becomes sub-relativistic, the flow approaches the non-relativistic Sedov-von Neumann-Talyor self-similar flow \citep{Sedov_1946,Taylor_1950}.

This section is organized as follows. The phase of rising flux, $t<t_{\text{peak}}$, is discussed in \S~\ref{sec:rising}, the asymptotic Sedov-von Neumann-Talyor phase, $t>t_{\text{ST}}$ is discussed in \S~\ref{sec:sedovtaylor}, and the intermediate phase between them is discussed in \S~\ref{sec:inter}. A summary of the analytic results is given in \S~\ref{sec:results}.

The hydrodynamic profiles are determined by three dimensional parameters, $\{M_0,n, c\}$ (the mass of the fast tail, the ISM density and the speed of light, respectively), and three dimensionless parameters, $\{\beta_0,s_{\rm ft},s_{\rm KN}\}$. The dependence of the profiles on the dimensional parameters follows directly from dimensional considerations. The flow depends non-trivially only on the dimensionless parameters. The approximations that we provide are valid over a wide range of values of these parameters, $5<s_{\rm ft}<12,~1<s_{\rm KN}<3,~0.3<\beta_0<0.9$.

\subsection{Rising phase}
\label{sec:rising}

In \S~\ref{sec:dynamics} we first derive an approximate description of the dependence of the velocity of the shocked plasma, $\beta_{\rm SP}$, and of the velocity $\beta_{\rm ej,RS}$ of the un-shocked plasma at the position reached by the reverse shock, on radius $R$. We then derive an approximate expression for $t_{\rm peak}$, the (observed) time at which the flux peaks. In \S~\ref{sec:ntf} we use these results to derive an approximate description of the non-thermal flux.

\subsubsection{Dynamics}
\label{sec:dynamics}
\begin{figure}
    \includegraphics[width=\columnwidth]{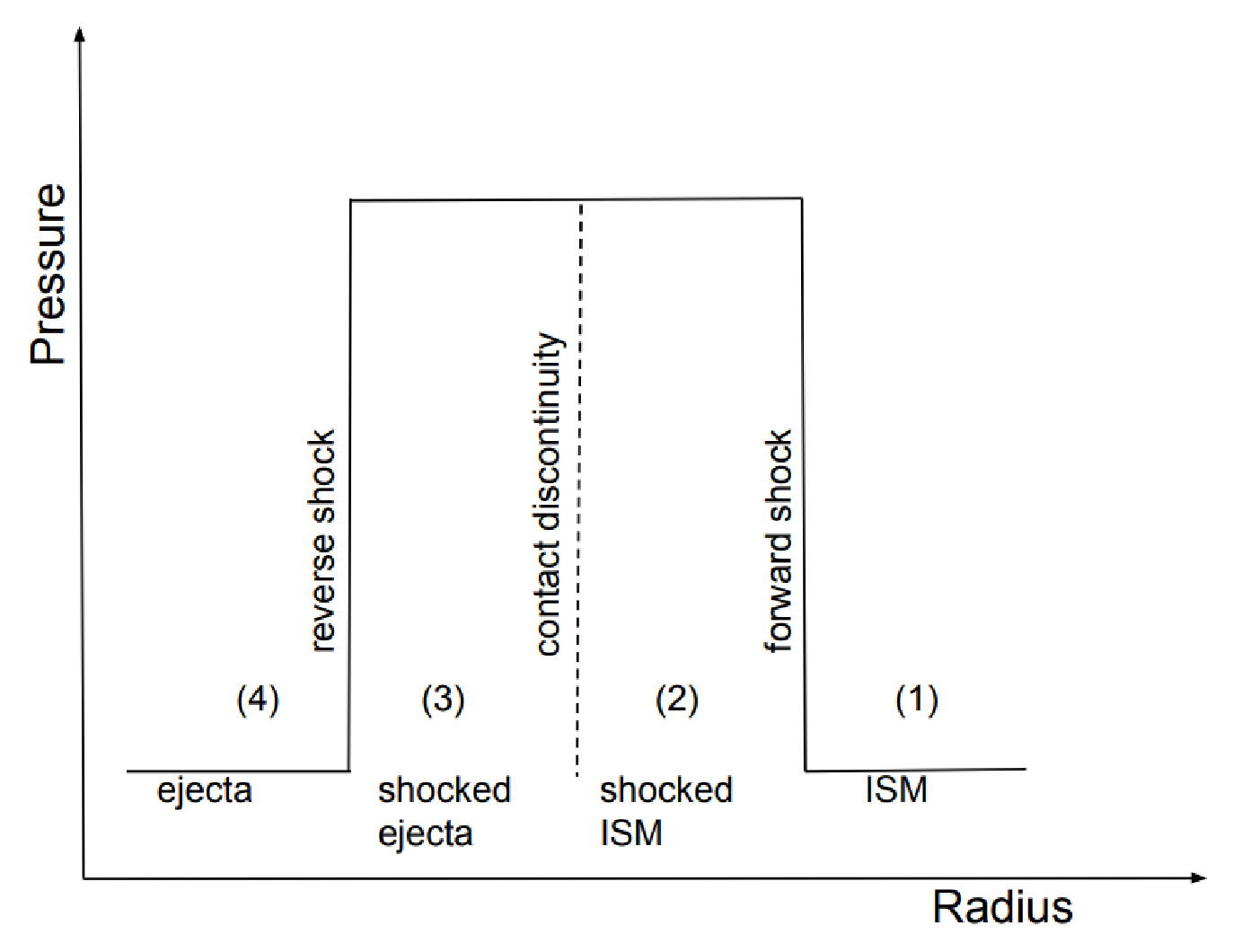}
    \caption{A schematic illustration showing the four regions of the forward-reverse shock structure described in \S~{\ref{sec:rising}}: (1) un-shocked ISM, (2) shocked ISM, (3) shocked ejecta, (4) un-shocked ejecta. The pressure in the unshocked regions is negligible compared to that in the shocked regions, $p_1,p_4\ll p_2,p_3$. The pressure and velocity are approximated in our analytic calculations as uniform within the shocked region, and the density is approximated as uniform within the two regions separated by the contact discontinuity.}
    \label{fig:FRshocks}
\end{figure}

Let us first obtain a qualitative estimate of the relation between the accumulated shocked ISM mass and the shocked ejecta mass. Consider ejecta of mass $M_{\rm ej}$ with uniform initial velocity, $\beta_{\rm ej}$. A significant deceleration of the shell occurs when a significant fraction of its energy is transferred to the ISM. Assuming that the velocity of the shocked plasma is not far below the initial shell velocity, i.e. that the deceleration by the reverse shock is modest as justified below, this implies significant deceleration when the shocked ISM mass, $M_{\rm ISM}$, reaches $M_{\rm ISM}\approx M_{\rm ej}/\gamma_{\rm ej}$. This is due to the fact that the shell's initial energy is $0.5\beta_{\rm ej}^2M_{\rm ej}c^2$ for $\beta_{\rm ej}\ll1$ and $\gamma_{\rm ej}M_{\rm ej}c^2$ for $\gamma_{\rm ej}\gg1$, while the shocked ISM energy is $\approx \beta_{\rm ej}^2M_{\rm ISM}c^2$ and $\approx \gamma_{\rm ej}^2M_{\rm ISM}c^2$ respectively. This result suggests that the position of the reverse shock within the ejecta, i.e. the initial velocity $\beta_{\rm ej,RS}$ of the ejecta shell that is reached by the reverse shock, is (implicitly) determined by $M(>\gamma_{\rm ej,RS}\beta_{\rm ej,RS})=\gamma_{\rm ej,RS}(R)M_{\rm ISM}(R)$, where $\gamma(R)$ is the Lorenz factor corresponding to $\beta(R)$.

A more accurate approximation, in particular taking into account the deceleration due to the reverse shock, is obtained as follows. We assume that at any given $R$ the velocity and pressure profiles within the shocked ejecta/ISM plasmas (see fig.~\ref{fig:FRshocks}) are uniform, and that the shocked ejecta/ISM densities are uniform within the regions separated by the contact discontinuity. Denoting the mass of the ejecta shocked by the reverse shock as $M_{\rm ej,RS}(R)$, we obtain (using Eq.~(\ref{eq:profile})) the un-shocked ejecta number density and velocity ahead of the reverse shock, $n_{\rm ej,RS}(R,M_{\rm ej,RS}(R))$ and $\beta_{\rm ej,RS}(R,M_{\rm ej,RS}(R))$. Given $n_{\rm ej,RS}(R,M_{\rm ej,RS}(R))$ and the number density of the ISM, $n$, a relation between $\beta_{\rm ej,RS}$ and the velocity of the shocked plasma, $\beta_{\rm SP}$, is obtained by requiring the pressures behind the forward and reverse shocks to be equal, thus determining $\beta_{\rm SP}(R,M_{\rm ej,RS}(R))$. Finally, $M_{\rm ej,RS}(R)$ is determined by requiring the energy in the shocked ejecta and ISM plasma to equal the initial kinetic energy carried by the ejecta mass $M_{\rm ej,RS}(R)$ (the thicknesses of the two shocked plasma regions, which are required for determining the internal energy of the shocked plasma, are determined by conservation of mass).

The solution for $\beta_{\rm SP}(R)$ is obtained by numerically solving the pressure equilibrium and energy conservation equations described in the preceding paragraph. We use an EoS of the form $p=(\hat{\gamma}-1)e$ with adiabatic index $\hat{\gamma}(e/n)$ varying from $4/3$ to $5/3$ following the analysis of \cite{Synge_1957} for a plasma of protons and electrons. This is not accurate for the shocked ejecta plasma, which is composed of a wide range of nuclei. However, the dependence of the results on the exact form of $\hat{\gamma}(e/n)$ for the shocked ejecta plasma is weak (as verified by the numeric calculations). Once the hydrodynamic evolution is determined, we calculate the non-thermal flux as described in the following subsection.

For the purpose of deriving analytic expressions for the flux, it is useful to derive analytic approximations for the dynamics. We give below several useful approximations, which we found to provide a good description of the dynamics obtained by solving the equations described above for a wide range of ejecta parameters, $5<s_{\rm ft}<12,~0.2<\gamma_{\rm ej,RS}\beta_{\rm ej,RS}<6$. We find that $\beta_{\rm ej,RS}(R)$ is well approximated by the implicit relation
\begin{equation}\label{eq:betaR}
  M(>\gamma_{\rm ej,RS}(R)\beta_{\rm ej,RS}(R))=\gamma_{\rm ej,RS}(R)M_{\rm ISM}(R),
\end{equation}
that the ratio of the pre- and post-shock velocities is approximately given by
\begin{equation}\label{eq:RS}
  \frac{\gamma_{\rm ej,RS}(R)\beta_{\rm ej,RS}(R)}{\gamma_{\rm SP}(R)\beta_{\rm SP}(R)}=1.7,
\end{equation}
and that $R\approx\beta_{\rm ej,RS}ct_{\rm s}$, where $t_{\rm s}$ is the time measured at the rest frame of the source ejecting the expanding plasma.

Let us consider next the radius and velocity of the shocked plasma dominating the emission that contributes to the flux observed by a distant observer at that time $t$. Photons emitted from the shock front along a line making an angle $\theta$ with respect to the line of sight are observed at time $t$ related to $R$ by  $R=\beta_{\rm ej,RS} ct/(1-\beta_{\rm ej,RS}\cos\theta)$ (where we have used $R\approx\beta_{\rm ej,RS}ct_{\rm s}$ and $t_s=t/(1-\beta_{\rm ej,RS}\cos\theta)$). The emission from a ring of opening angle $\theta$ and thickness $d\theta$ increases with $\theta$ due to the increase in ring area, and decreases with $\theta$ due to the decrease in the Lorenz boost, $1/(\gamma^2(1-\beta\cos\theta)^2)$. We therefore estimate the angle dominating the flux observed at time $t$ as the angle for which $\partial_\theta( 2\pi \sin\theta/(\gamma^2(1-\beta\cos\theta)^2))=0$,
\begin{equation}
   \cos\theta= -\frac{1}{2\beta}+\sqrt{2+\frac{1}{4\beta^2}}.
\end{equation}
We therefore approximate
\begin{equation}
    R=\frac{\beta_{\rm ej,RS} ct}{\left(1.5-\sqrt{2\beta_{\rm ej,RS}^2+0.25}\right)}.
\end{equation}
Adopting this approximation we have
\begin{equation}
\begin{aligned}
\label{eq:gMt}
  \gamma_{\rm ej,RS}M_{\rm ISM}(t) &= \frac{4\pi}{3}n m_pc^3 \frac{\gamma_{\rm ej,RS}\beta_{\rm ej,RS}^3}{\left(1.5-\sqrt{2\beta_{\rm ej,RS}^2+0.25}\right)^3}t^3 \\
  &\approx   \frac{4\pi}{3}n m_pc^3 12(\gamma_{\rm ej,RS}\beta_{\rm ej,RS})^{5.5}t^3,
  \end{aligned}
\end{equation}
where the last approximation holds (to $\sim10$\%) for $0.3<\gamma\beta<2.5$. Using Eq.~(\ref{eq:betaR}) we finally obtain
\begin{equation}
\label{eq:tR}
    \gamma_{\rm ej,RS}\beta_{\rm ej,RS}=\left(\frac{t}{t_R}\right)^{-\frac{3}{5.5+s_\text{ft}}},\quad
    t_R\equiv \left(\frac{M_0(\gamma_0\beta_0)^{s_\text{ft}}}{16 \pi nm_pc^3}\right)^\frac{1}{3}.
\end{equation}
Eqs.~(\ref{eq:RS}) and~(\ref{eq:tR}) determine the velocity of the shocked plasma dominating the emission of flux measured by a distant observer at time $t$.

Using the above approximations, we find that the observed time at which the flux peaks, i.e. the observed time at which the reverse shock reaches the shallow $\gamma\beta<\gamma_0\beta_0$ part of the ejecta, is given by
\begin{equation}
\label{eq:tpeak}
\begin{aligned}
      t_{\text{peak}}&\approx \left(\frac{3M_0}{4\pi nm_p\gamma_{0}}\right)^{\frac{1}{3}} \left(\frac{1.5-\sqrt{0.25+2\beta_0^2}}{c\beta_{0}}\right).
\end{aligned}
\end{equation}

\subsubsection{Non-thermal flux}
\label{sec:ntf}

The solution obtained for $\beta_{\rm SP}(R)$ (and $\beta_{\rm FS}(R)$, $\beta_{\rm ej,RS}(R)$) by numerically solving the pressure equilibrium and energy conservation equations described in the preceding subsection, provide, under the approximation of uniform pressure, velocity, and density within the shocked plasma region, a full approximate description of the hydrodynamics- $p(r,t)$, $v(r,t)$, and $n(r,t)$. We use the hydrodynamic profiles to calculate the specific emissivity at the plasma frame, $j'_\nu(r,t)$ \citep[following][]{Rybicki_1979}, using the phenomenological description adopted for the magnetic field and for the electron energy distribution- assuming that the magnetic field and electrons carry fractions $\varepsilon_B$ and $\varepsilon_e$ of the internal energy density respectively, and that the electrons follow a power-law energy distribution. Using $j_\nu$, the specific intensity $I_\nu(\hat{\Omega})$ and the specific flux, $F_\nu$, are derived numerically (properly accounting for relativistic and arrival time effects).

We derive below analytic expressions for $F_\nu$, which provide a good (to 10's of percent) approximation for $F_\nu$ obtained numerically as described above for a wide range of parameters, $5<s_{\rm ft}<12,~0.2<\gamma_{\rm ej,RS}\beta_{\rm ej,RS}<6$. We consider in this derivation only the emission from the shocked ISM, that dominates over the emission from the shocked ejecta (as verified by the numeric calculations). The derivation given below provides the dependence on model parameters. Order unity corrections are applied to the numeric coefficients obtained in the derivation, in order to improve the agreement with the numeric results.
We do not show in this section comparisons of the approximations to the numeric results. Instead, we show in \S~\ref{sec:numeric} direct comparison to the full numeric calculation, which includes a full numeric solution of the hydrodynamics equations (instead of the approximate treatment used in this subsection).

We first consider the flux produced by a spherical shock driven by a piston of fixed Lorenz factor $\gamma$, and then derive the flux obtained for the ejecta under consideration by using the approximate analytic dependence of $\gamma(t)$ derived in the preceding subsection. Approximating the internal energy carried by the shocked plasma as $E(R)=\gamma(\gamma-1)M(R)c^2$, the luminosity observed by a distant observer, for the case where electrons lose all their energy via synchrotron emission on a time scale short compared to the expansion time, is $L=\varepsilon_e dE/dt$. We denote by $\gamma_c$ the (plasma frame) electron Lorenz factor above which the electron energy loss time is short (compared to the expansion time), and by $\nu_c$ the corresponding (observer frame) synchrotron frequency. Noting that for a power-law energy distribution of electrons the specific flux at $\nu>\nu_c$ follows $F_\nu\propto\nu^{-p/2}$, we have for $\nu>\nu_c$
\begin{equation}
\label{eq:Lcnu}
    L_\nu=2\pi\gamma(\gamma-1)\frac{R^3}{t} \varepsilon_e n m_p c^2
    l_p \frac{\nu^{-p/2}}{\nu_m^{1-p/2}},
\end{equation}
where
\begin{equation}\label{eq:lp}
  l_p=\frac{p-2}{1-(\gamma_\text{max}/\gamma_\text{min})^{2-p}},\quad
  l_p\xrightarrow[p \to 2]{}\frac{1}{\ln(\gamma_\text{max}/\gamma_\text{min})},
\end{equation}
and $\nu_m$ is the characteristic observed frequency of the synchrotron emission of the lowest energy electrons, with $\gamma_e=\gamma_{\rm min}$. Based on particle acceleration calculations as well as observations, we expect electrons to be accelerated over many decades in energy, $\gamma_\text{max}/\gamma_\text{min}\gg1$. The dependence on $\gamma_\text{max}/\gamma_\text{min}$ is thus weak. When presenting numeric results below, we use $\gamma_\text{max}/\gamma_\text{min}=10^5$ \citep{Sironi_2013}.

For $\nu<\nu_c$, the specific luminosity is suppressed by a factor $(\nu/\nu_c)^{1/2}$. The self-absorption frequency, $\nu_a$, is expected to be low, and we therefore consider the emission only at $\nu>\nu_a$. 

Let us estimate next the frequencies $\nu_m$ and $\nu_c$. The characteristic plasma frame frequency of the synchrotron emission of electrons with Lorenz factor $\gamma_e$ and isotropic velocity distribution is $\gamma_e^2\sqrt{(\pi u_B/2)}e/(m_ec)$, where $u_B$ is the (plasma frame) magnetic field energy density. Approximating the post shock energy density, $2nm_p v^2$ for non-relativistic and $4\gamma^2nm_p c^2$ for highly relativistic shocks, by $4\gamma(\gamma-1)nm_p c^2$, we have $u_B=4\varepsilon_B\gamma(\gamma-1)n m_p c^2$. Defining
\begin{equation}\label{eq:nus}
  \nu_s=\frac{e}{m_e}\sqrt{2\pi\varepsilon_B n m_p},
\end{equation}
the observer frame synchrotron frequencies are thus
\begin{equation}\label{eq:freq}
  \nu_m=\frac{\gamma_{\rm min}^2\gamma^{-\frac{1}{2}}(\gamma-1)^{\frac{1}{2}}\nu_s}{\left(1.5-\sqrt{2\beta^2+0.25}\right)},\quad \nu_c=\frac{\gamma_c^2\gamma^{-\frac{1}{2}}(\gamma-1)^{\frac{1}{2}}\nu_s}{\left(1.5-\sqrt{2\beta^2+0.25}\right)}.
\end{equation}
$\gamma_{\rm min}$ is obtained by requiring the average energy per electron to equal a fraction $\varepsilon_e$ of the post-shock internal energy per particle, $(\gamma-1)m_pc^2$, while $\gamma_c$ is obtained by requiring the (plasma frame) synchrotron loss time, $m_ec^2/(\gamma_e(4/3) \sigma_T cu_B)$, to be equal to the time measured at the plasma frame, $t/\gamma/\left(1.5-\sqrt{2\beta^2+0.25}\right)$. This yields
\begin{equation}
\begin{aligned}
\label{eq:gs}
  \gamma_{\rm min}&=\frac{l_p(\gamma-1)}{p-1}\frac{\varepsilon_em_p}{m_e},\\
  \gamma_c&=
  \frac{\left(1.5-\sqrt{2\beta^2+0.25}\right)m_e}{4(\gamma-1)\varepsilon_B\left(\frac{4}{3}\right)\sigma_T n m_p ct}.
\end{aligned}
\end{equation}

Eqs.~(\ref{eq:Lcnu}-\ref{eq:gs}) provide an approximate description of $L_\nu(t)$ for time independent $\gamma$. In order to obtain an approximate analytic description of $L_\nu$ for the shock driven by the ejecta described by Eq.~(\ref{eq:profile}), we replace the time independent $\beta$ and $R(t)$ in these equations with $\beta_{\rm SP}(t)$ and $R(t)$ obtained in the preceding sub-section, using the appropriate time derivative $dE/dt$. We approximate the resulting functional dependence on $\gamma$ and $\beta$ by power-laws of the form $\propto(\gamma\beta)^\lambda$. This enables us to use Eqs.~(\ref{eq:RS}) and~(\ref{eq:tR}) to derive analytic expressions for $\nu_c$, $\nu_m$ and $L_\nu$. For $\nu_m$, for example, we have
\begin{equation}
\label{eq:num_aprx}
\begin{aligned}
  \nu_m&=\left(\frac{l_p(\gamma-1)}{p-1}\frac{\varepsilon_em_p}{m_e}\right)^2\frac{\gamma^{-1/2}(\gamma-1)^{\frac{1}{2}}\nu_s}{\left(1.5-\sqrt{2\beta^2+0.25}\right)}\\
  &\approx\left(\frac{l_p\varepsilon_em_p}{(p-1)m_e}\right)^2 0.2(\gamma_{\rm SP}\beta_{\rm SP})^{5.1}\nu_s,
  \end{aligned}
\end{equation}
where the approximation holds (to 10\%) for $0.3<\gamma_{\rm SP}\beta_{\rm SP}<2.5$. Using this approximation with Eqs.~(\ref{eq:RS}) and~(\ref{eq:tR}) yields an analytic expression for $\nu_m(t)$. The $(\gamma\beta)^\lambda$ power-law approximations used for $\nu_c$ and $L_\nu$ are given in App. \ref{appendix_power-law}, and the resulting analytic expressions are given in \S~\ref{sec:results}.

\subsection{Sedov–von Neumann–Taylor (SvNT) phase}
\label{sec:sedovtaylor}

At late time, after the reverse shock crosses the entire ejecta and once the ejecta decelerates to sub-relativistic, $\beta_\text{s}\lesssim0.05$, velocity, the flow approaches the self-similar behavior described by the SvNT solution for an explosion with energy given by the initial kinetic energy, $E$, of the ejecta. At this stage, the shock radius is given by $R_\text{ST}(t)=\zeta\left(\frac{E}{\rho}\right)^\frac{1}{5}t^\frac{2}{5}$, where $\zeta$ is an order unity dimensionless parameter depending on the adiabatic index $\hat{\gamma}$ of the shocked material, $\zeta= 1.15$ for $\hat{\gamma}=\frac{5}{3}$. We therefore estimate the time beyond which the flow is well approximated by the SvNT solution as
\begin{equation}
\label{eq:t_ST}
   t_{\text{ST}}\approx\left(0.05 \frac{2.5c}{\zeta}\left(\frac{E}{\rho}\right)^{-\frac{1}{5}}\right)^{-\frac{5}{3}},
\end{equation}

The synchrotron luminosity produced by the collisionless shock is derived in a manner similar to that of the preceding sub-section. Assuming that a fraction $\varepsilon_e$ of the internal energy flux through the shock, $4\pi\frac{4\hat{\gamma}}{(\hat{\gamma}+1)^2}\frac{1}{2}\rho v_s^3R^2$, is carried by electrons, the luminosity at frequencies $\nu>\nu_c$ (where electrons lose all their energy over a time short compared to the expansion time), is given by 
\begin{equation}
\label{eq:L_ST}
L_{\nu,\text{ST}}= \frac{4\pi\hat{\gamma}}{(\hat{\gamma}+1)^2}\varepsilon_e\rho v_s^3R^2l_p\frac{\nu^{-\frac{p}{2}}}{\left(\beta^2_\text{min}\nu_{m,\text{ST}}\right)^{1-\frac{p}{2}}},
\end{equation}
where
\begin{equation}
\label{eq:b_ST}
    \beta_\text{min}= \frac{l_p}{p-1}\varepsilon_e\frac{m_p}{m_e}\frac{9\beta_s^2}{32}
\end{equation}
is the velocity (at the plasma frame) of the lowest energy electrons. Note that the lowest energy electrons are not relativistic for $\beta_s\ll1$, and we have therfore used in this calculation an electron energy distribution that follows a power-law in momentum, $dn_e/d(\gamma\beta)_e\propto (\gamma\beta)_e^{-p}$ \citep{Krymskii_1977,Blandford_1978,Bell_1978}. 
The corresponding observer frame minimum and cooling frequencies are 
\begin{equation}\label{eq:freqST}
  \nu_{m,\text{ST}}=\frac{3\beta_s}{4\sqrt{2}}\nu_s,\quad \nu_{c,\text{ST}}=\gamma_{c,\text{ST}}^2\frac{3\beta_s}{4\sqrt{2}}\nu_s,
\end{equation}
where 
\begin{equation}
\label{eq:gST}
    \gamma_{c,\text{ST}}=\frac{8m_e}{9\beta_s^2\varepsilon_B\left(\frac{4}{3}\right)\sigma_T\rho ct}.
\end{equation}

\begin{figure}
    \centering
    \includegraphics[width=\columnwidth]{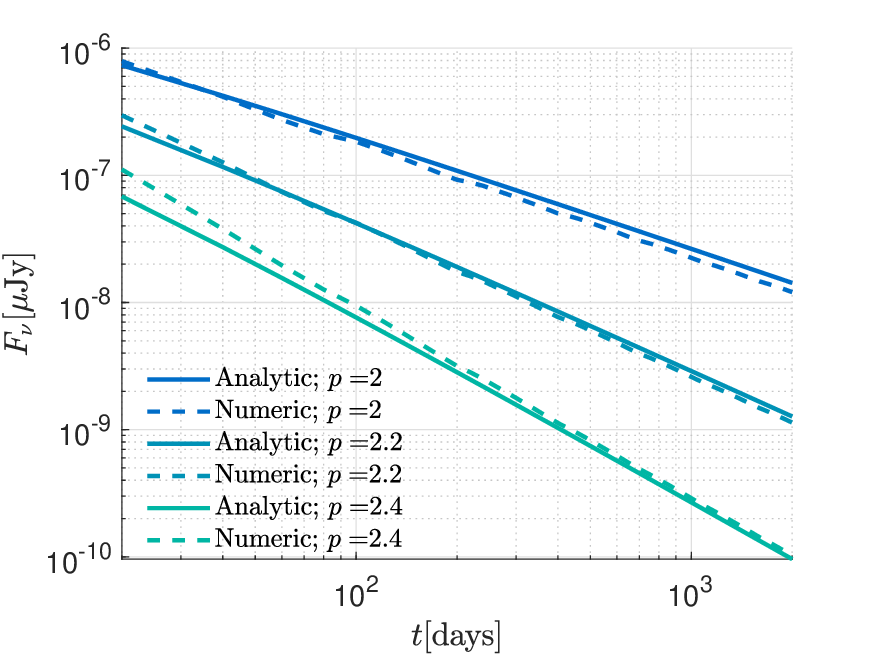}
    \caption{A comparison of the analytic (solid lines, Eqs.~(\ref{eq:ST}-\ref{eq:ST2})) and numeric (dashed lines) calculations of the synchrotron flux during the phase of non-relativistic self-similar expansion (where the flow is described by the SvNT solutions). The parameter values chosen for the calculations are $E=10^{42}$erg, $\varepsilon_e=10^{-1},\varepsilon_B=10^{-2}$ and $n=10^{-2}{\rm cm}^{-3}$. Results are shown for $\nu_{m,\text{ST}},\nu_{a,\text{ST}}<\nu=3\text{GHz}<\nu_{c,\text{ST}}$ and observer distance of $D=10$Mpc.}
    \label{fig:flux_sedov}
\end{figure}

The specific flux obtained using Eqs.~(\ref{eq:L_ST})-(\ref{eq:gST}) is given below by Eqs.~(\ref{eq:ST}-\ref{eq:ST2}). The analytic description provides a good approximation (within 10's of percent) to the results obtained by numerically calculating the synchrotron flux using the analytic self-similar SvNT hydrodynamic profiles (calculating $j_\nu$, $I_\nu$ and $F_\nu$ adopting the phenomenological assumptions described above regarding the magnetic field energy density and the electron energy distribution). This is demonstrated in figure~\ref{fig:flux_sedov}, showing a comparison of the analytic and numeric results for $\nu_{m,\text{ST}},\nu_{a,\text{ST}}<\nu<\nu_{c,\text{ST}}$.\newline

\subsection{Intermediate phase}
\label{sec:inter}

We consider here the phase during which the reverse shock propagates through the shallower, $\beta<\beta_0$, part of the ejecta. Using the semi-analytic calculation of the dynamics described in \S~\ref{sec:dynamics}, we find that $\gamma_{\rm ej,RS}\beta_{\rm ej,RS}$ is well described for $1<s_{\rm KN}<3$ and $0.1<\gamma_{\rm ej,RS}\beta_{\rm ej,RS}<0.7$ by
\begin{equation}
\label{eq:inter}
\gamma_{\rm ej,RS}\beta_{\rm ej,RS}\propto t^{-3/(4.7+s_{\rm KN})}.
\end{equation}
The synchrotron flux is then derived in a manner similar to that of \S~{\ref{sec:ntf}}. The $(\gamma\beta)^\lambda$ power-law approximations used for $\nu_c$ and $L_\nu$ in this regime are given in App.~\ref{appendix_power-law}, and the resulting analytic expressions are given in \S~\ref{sec:results}.

\subsection{Summary of the analytic model results}
\label{sec:results}

We summarize below the analytic results derived in the preceding sub-sections for the specific flux at the three time regimes, $t<t_{\rm peak}$, $t>t_{\rm ST}$ and $t_{\rm peak}<t<t_{\rm ST}$, and provide at the end an interpolation formula that provides a good approximation for the specific flux at all times. The numerical values of the parameters are normalized to those expected to be relevant for NS merger ejecta.

Analytic expressions for the synchrotron emission produced by an ejecta with a power-law dependence of ejecta energy on momentum, Eq.~(\ref{eq:profileE}), may be obtained from the results given in this sub-section by the substitutions
\begin{equation}
\begin{aligned}
\label{eq:Eofv}
  M_0&=1.5\frac{E_0}{(\gamma_0\beta_0)^{2}c^2},\\
  s_{\rm ft}&=\alpha_{\rm ft}+2,\quad s_\text{KN}=\alpha_\text{KN}+1.5.
  \end{aligned}
\end{equation}

\begin{figure}
    \centering
    \includegraphics[width=\columnwidth]{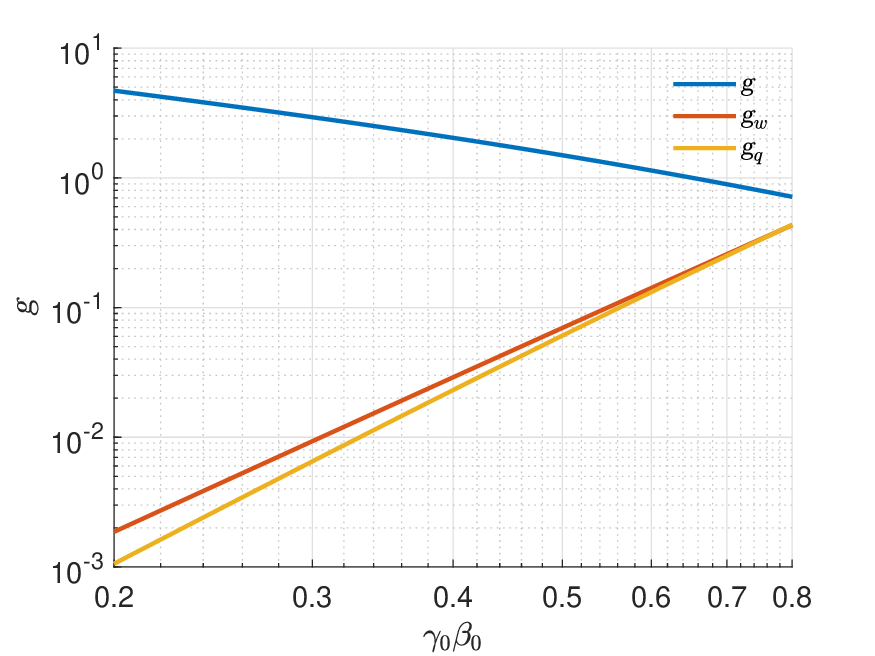}
    \caption{The functions $g(\beta_0)$ and $g_{w,q}(\beta_0,p,s)$ (Eqs. (\ref{eq:times}) and~(\ref{eq:gfactor})) for $s=7$, $p=2.2$.}
    \label{fig:g}
\end{figure}

The three characteristic time scales are given by Eqs.~(\ref{eq:tR}),~(\ref{eq:tpeak}) and~(\ref{eq:t_ST})
\begin{equation}
\label{eq:times}
\begin{aligned}
  t_R&=51\left(\frac{M_{R,-6}}{n_{-2}}\right)^{\frac{1}{3}}\text{days},\\
  t_\text{peak}&= 550g(\beta_0)\left(\frac{M_{0,-4}}{n_{-2}}\right)^{\frac{1}{3}}\text{days},\quad
   g(\beta_0)=\frac{1.5-\sqrt{0.25+2\beta_0^2}}{\gamma_0^{\frac{1}{3}}\beta_{0}},\\
  t_\text{ST}&=2.9\times10^4\left(\frac{E_{50}}{n_{-2}}\right)^{\frac{1}{3}}\text{days},
  \end{aligned}
\end{equation}
where
\begin{equation}\label{eq:M0}
  M_R\equiv M(\gamma\beta>1)= M_0(\gamma_0\beta_0)^{s_{\rm ft}},
\end{equation}
and we use $n=10^{-2}n_{-2}{\rm cm}^{-3}$, $E=10^{50}E_{50}$erg (where $E$ is the total kinetic energy of the ejecta, $E(\gamma\beta>0.1)$), $M_R=10^{-6}M_{R,-6}M_\odot$ and $M_0=10^{-4}M_{0,-4}M_\odot$. The function $g(\beta_0)$ is plotted in Fig.~\ref{fig:g}.
For an ejecta with a power-law dependence of ejecta energy on momentum, Eq.~(\ref{eq:profileE}),
\begin{equation}\label{eq:E0}
  E_R\equiv E(\gamma\beta>1)= E_0(\gamma_0\beta_0)^\alpha =\frac{M_Rc^2}{1.5}.
\end{equation}
The specific flux at $t<t_\text{peak}$ is given by Eqs.~(\ref{eq:Lcnu}-\ref{eq:gs})
\begin{equation}
\begin{aligned}
\label{eq:an_end1}
    F_{\nu,\rm ft}&=D_{26.5}^{-2}\varepsilon_{e,-1}^{p-1}f_{\rm ft}(p) \\ &\times\begin{cases}
    7.7\varepsilon_{B,-2}^{\frac{p-2}{4}}n_{-2}^{\frac{3p-2}{4}}M_{R,-6}^\frac{2}{3}\nu_{9.5}^{-\frac{p}{2}}\left(\frac{t}{t_R}\right)^{w_{\rm ft}} m\text{Jy} & \text{for}\quad \nu>\nu_{c,\rm ft},\\
    0.1\varepsilon_{B,-2}^{\frac{p+1}{4}}n_{-2}^{\frac{p+1}{4}}M_{R,-6}\nu_{9.5}^\frac{1-p}{2}\left(\frac{t}{t_R}\right)^{q_{\rm ft}}\mu\text{Jy} & \text{for}\quad \nu<\nu_{c,\rm ft},
    \end{cases}
\end{aligned}
\end{equation}
with $D=10^{26.5}D_{26.5}$cm, $\nu=10^{9.5}\nu_{9.5}$Hz, $\varepsilon_{e}=10^{-1}\varepsilon_{e,-1}$ and $\varepsilon_{B}=10^{-2}\varepsilon_{B,-2}$,
\begin{equation}
\label{eq:an_end2}
w_{\rm ft}=\frac{5-7.5p+2s_{\rm ft}}{5.5+s_{\rm ft}},\quad q_{\rm ft}=\frac{4.5-7.5p+3s_{\rm ft}}{5.5+s_{\rm ft}},
\end{equation}
and
\begin{equation}
    f_{\rm ft}(p)=960\times8.8^{-p}(2.5-0.7p)(p-1)^{2-p}(l_p)^{p-1}.
\end{equation}
$f_{\rm ft}(p)$ deviates from unity by less than $30\%$ for $2<p<2.5$ (the origin of $8.8^{-p}(2.5-0.7p)$ factor is in the normalization chosen for the parameter values and the power-law approximation (App. \ref{appendix_power-law})).

The cooling frequency and $\gamma_{\rm SP}\beta_{\rm SP}(t)$ at $t<t_\text{peak}$ are given by Eqs.~(\ref{eq:RS}),~(\ref{eq:tR}) and (\ref{eq:freq})
\begin{equation}
\begin{aligned}
\label{eq:an_end3}
         \nu_{c,\rm ft}&= 1.9\times 10^{10}\varepsilon_{B,-2}^{-\frac{3}{2}}n_{-2}^{-\frac{5}{6}}M_{R,-6}^{-\frac{2}{3}}\left(\frac{t}{t_R}\right)^{\frac{0.7-2s_{\rm ft}}{5.5+s_{\rm ft}}}\text{GHz},\\
         \gamma_{\rm SP}\beta_{\rm SP}&=0.6\left(\frac{t }{t_R}\right)^{-\frac{3}{5.5+s}}.\\
\end{aligned}
\end{equation}
Notice that in the limit $s_{\rm ft}\rightarrow \infty$, in which the mass increases very rapidly as the velocity is reduced, we have $\gamma\beta\propto t^0$ and $L_\nu\propto t^{3(2)}$ below (above) $\nu_c$, corresponding to no deceleration.

The peak flux, $F_{\nu,\rm ft}(t_\text{peak})$, is given by
\begin{equation}
    \begin{aligned}
\label{eq:Fpeak}
    F_{\nu,\rm peak}&=D_{26.5}^{-2}\varepsilon_{e,-1}^{p-1}f_{\rm ft}(p) \\ &\times\begin{cases}
    170\varepsilon_{B,-2}^{\frac{p-2}{4}}n_{-2}^{\frac{3p-2}{4}}M_{0,-4}^\frac{2}{3}\nu_{9.5}^{-\frac{p}{2}}g_w m\text{Jy} & \text{for}\quad \nu>\nu_{c,\rm peak},\\
    10\varepsilon_{B,-2}^{\frac{p+1}{4}}n_{-2}^{\frac{p+1}{4}}M_{0,-4}\nu_{9.5}^\frac{1-p}{2}g_q \mu\text{Jy} & \text{for}\quad \nu<\nu_{c,\rm peak},
    \end{cases}
    \end{aligned}
\end{equation}
where $\nu_{c,\rm peak}=\nu_{c,ft}(t_{\rm peak})$ and
\begin{equation}
\begin{aligned}
\label{eq:gfactor}
g_w&=2.3^{w_{\rm ft}}(\gamma_0\beta_0)^{\frac{s_{\rm ft}}{3}(2-w_{\rm ft})}(g(\beta_0))^{w_{\rm ft}}\approx1.1(\gamma_0\beta_0)^{2.3p-1},\\
g_q&=2.3^{q_{\rm ft}} (\gamma_0\beta_0)^{s_{\rm ft}\left(1-\frac{q_{\rm ft}}{3}\right)} (g(\beta_0))^{q_{\rm ft}}\approx1.2(\gamma_0\beta_0)^{2.2p-0.5}.
\end{aligned}
\end{equation}
The functions $g_{w,q}(\beta_0,p=2.2,s_{\rm ft}=7)$ are plotted in Fig.~\ref{fig:g}.

The specific flux at $t>t_\text{ST}$ is given by Eqs. (\ref{eq:t_ST}-\ref{eq:gST})
\begin{equation}
\begin{aligned}
\label{eq:ST}
    F_{\nu,\text{ST}}&=D_{26.5}^{-2}\varepsilon_{e,-1}^{p-1}f_{\rm ST}(p) \\ &\times
    \begin{cases}
    1\varepsilon_{B,-2}^{\frac{p-2}{4}}n_{-2}^{\frac{3p-2}{4}}E_{50}^\frac{2}{3}\nu_{9.5}^{-\frac{p}{2}}\left(\frac{t}{t_\text{ST}}\right)^{w_\text{ST}} m\text{Jy} & \text{for}\quad \nu>\nu_{c,\text{ST}},\\
    0.1\varepsilon_{B,-2}^{\frac{p+1}{4}}n_{-2}^{\frac{p+1}{4}}E_{50}\nu_{9.5}^\frac{1-p}{2}\left(\frac{t}{t_\text{ST}}\right)^{q_\text{ST}}\mu\text{Jy} & \text{for}\quad \nu<\nu_{c,\text{ST}},
    \end{cases}
    \end{aligned}
\end{equation}
with
\begin{equation}
\label{eq:ST2}
w_\text{ST}=\frac{20-15p}{10},\quad
q_\text{ST}=\frac{21-15p}{10},
\end{equation}
and
\begin{equation}
    f_{\rm ST}(p)=1.4\times10^{10}\left(2\times10^4\right)^{-p}(p-1)^{2-p}(l_p)^{p-1}.
\end{equation}
$f_{\rm ST}(p)$ is bounded by $0.75<f_{\rm ST}(p)<3$ for $2<p<2.5$ (The origin of the $\left(2\times10^4\right)^{-p}$ factor is the normalization chosen for the values of the parameters).

The cooling frequency at $t>t_\text{ST}$ is given by Eqs. (\ref{eq:freqST}-\ref{eq:gST})
\begin{equation}
     \nu_{c,\text{ST}}= 3.7\times10^8\varepsilon_{B,-2}^{-\frac{3}{2}}E_{50}^{-\frac{2}{3}}n_{-2}^{-\frac{5}{6}}\left(\frac{t}{t_\text{ST}}\right)^{-\frac{1}{5}}\text{GHz}.\\
\end{equation}

For $t_\text{peak}<t<t_\text{ST}$ the specific flux is given by
\begin{equation}
\label{eq:an_end1_2}
    F_{\nu,\text{KN}}=F_{\nu,\rm peak}\times\begin{cases}\left(\frac{t}{t_\text{peak}}\right)^{w_{\rm KN}}\text{for}\quad \nu>\nu_c,\text{KN},
    \\\left(\frac{t}{t_\text{peak}}\right)^{q_{\rm KN}} \text{for}\quad \nu<\nu_c,\text{KN},
    \end{cases}
\end{equation}
with
\begin{equation}
\label{eq:an_end2_2}
w_{\rm KN}=\frac{7.4-7.5p+2s_{\rm KN}}{4.7+s_{\rm KN}},\quad q_{\rm KN}=\frac{7.5-7.5p+3s_{\rm KN}}{4.7+s_{\rm KN}}.
\end{equation}
The cooling frequency is given by
\begin{equation}
\label{eq:an_end3_2}
    \nu_{c,\text{KN}}=\nu_{c,\text{ft}}(t_{\rm peak}) \left(\frac{t}{t_{\rm peak}}\right)^{\frac{0.5-2s_{\rm ft}}{4.7+s_{\rm ft}}}.
\end{equation}

The $\nu<\nu_c$ specific flux is well described at all time by the interpolation
\begin{equation}
\label{eq:betw}
\begin{aligned}
&F_\nu=\\
&\left(0.5F_{\nu,\text{peak}}^{-5}\left(\left(\frac{t}{t_\text{peak}}\right)^{-5q_{\rm ft}}+\left(\frac{t}{t_\text{peak}}\right)^{-5q_\text{KN}}\right)+\left(F_{\nu,\text{ST}}(t_\text{ST})\left(\frac{t}{t_\text{ST}}\right)^{q_\text{ST}}\right)^{-5}\right)^{-0.2}.
\end{aligned}
\end{equation}
Here, $F_{\nu,\text{peak}}$ is given by Eq. (\ref{eq:Fpeak}), $t_\text{peak}$ is given by Eq. (\ref{eq:times}), $q_{\rm ft}$ is given by Eq. (\ref{eq:an_end2}), $q_\text{KN}$ is given by Eq. (\ref{eq:an_end2_2}) and $F_{\nu,\text{ST}}$ is given by Eq. (\ref{eq:ST}). The specific flux at $\nu>\nu_c$ is well described by a similar expression, replacing $q_X$ with $w_X$ given by Eqs. (\ref{eq:an_end2}), (\ref{eq:an_end2_2}) and (\ref{eq:ST2}).

\section{Comparison to numeric calculations}
\label{sec:numeric}

We compare in this section the approximate analytic description of the non-thermal specific flux derived in the preceding section with the results obtained using full numeric solutions of the special relativistic 1D hydrodynamics equations (note that these are solutions of the exact hydrodynamic equations, rather than of the approximate equations derived in \S~\ref{sec:dynamics}). The numeric code we use is a 1D Lagrangian code using a predictor-corrector scheme, artificial viscosity, and time steps set by Courant–Friedrichs–Lewy (CFL) conditions. The code was verified by testing it against standard test problems \citep{Sedov_1946,Marti_2003,Guzman_2012}.
The EoS used in the numeric calculations is the same as that used for the approximate hydrodynamic solutions described in \S~\ref{sec:dynamics}, $p=(\hat{\gamma}-1)e$ with adiabatic index $\hat{\gamma}(e/n)$ varying from $4/3$ to $5/3$ following the analysis of \cite{Synge_1957}.
The initial conditions used for the numerical calculations were: density and velocity profiles given by Eq. (\ref{eq:profile}) or (\ref{eq:profileE}) (the radius of a shell of mass $M(>\gamma\beta)$ is given by $r=\beta ct_0$ for some small chosen $t=t_0$), surrounded by a static uniform cold gas of number density $n$. The initial pressure is set to zero everywhere. 

\begin{figure}
 \centering
\begin{subfigure}[b]{0.5\textwidth}
   \centering
    \includegraphics[width=\columnwidth]{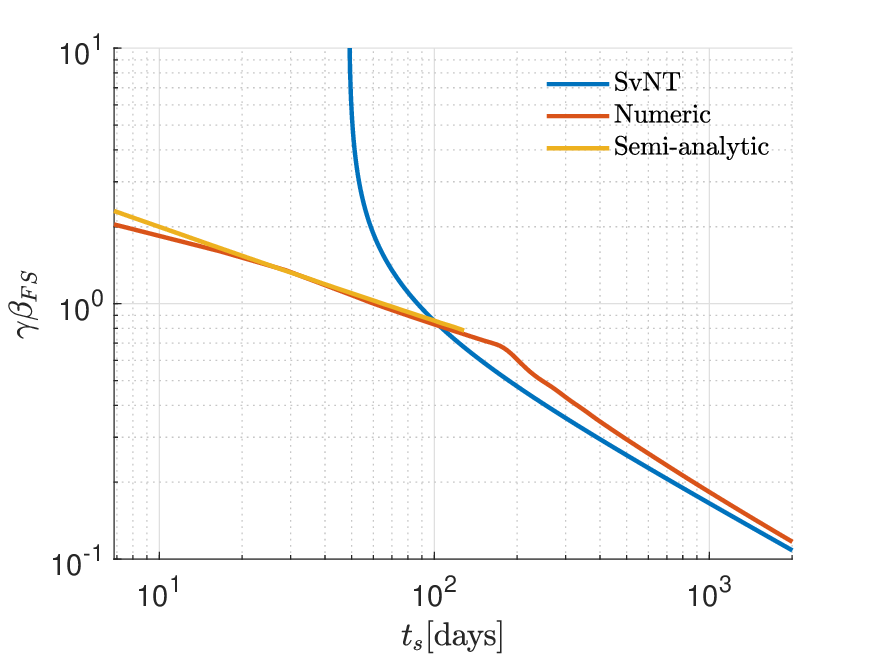}
\end{subfigure}
\begin{subfigure}[b]{0.5\textwidth}
    \centering
    \includegraphics[width=\columnwidth]{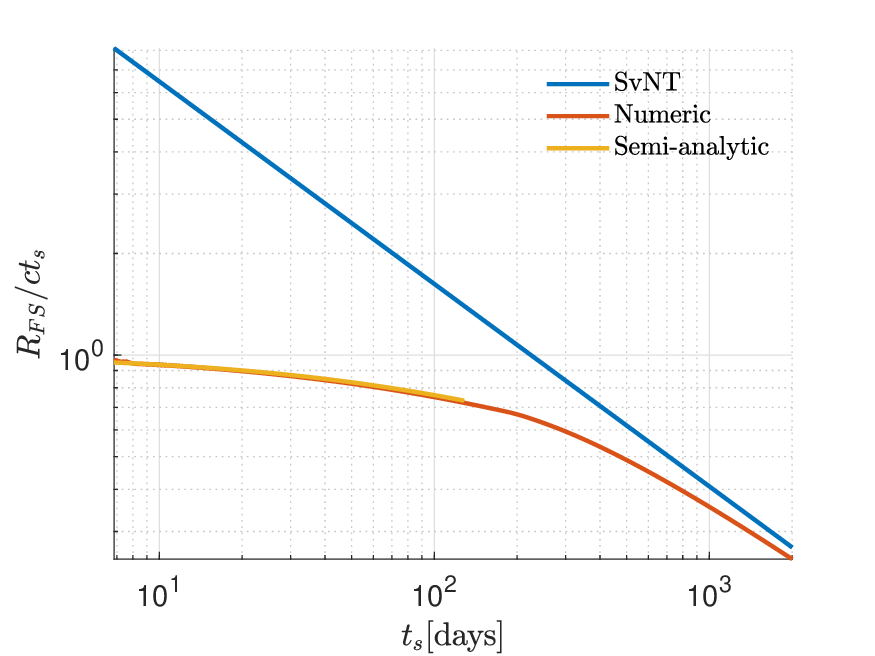}
    \end{subfigure}
    \caption{The forward shock $\gamma\beta$ (upper panel) and radius (lower panel) as a function of $t_{\rm s}$, the time measured at the rest frame of the source ejecting the expanding plasma. The results of the numeric solution of the hydrodynamics (red) are compared with the semi-analytic solution described in \S~\ref{sec:dynamics} (yellow), valid at early time (up to the time at which the reverse shock crosses the fast tail, $s_\text{ft}>5$, $\sim150$days), and with the SvNT self-similar solution (red) valid at late times. The parameter values used for this calculation are $n=3\times10^{-2}\text{cm}^{-3},M_0=2\times10^{-6}M_\odot,s_{\rm ft}=7,s_\text{KN}=1.5$ and $\gamma_0=1.35$.}
    \label{fig:compar1}
\end{figure}

\begin{figure}
 \centering
\begin{subfigure}[b]{0.5\textwidth}
   \centering
    \includegraphics[width=\columnwidth]{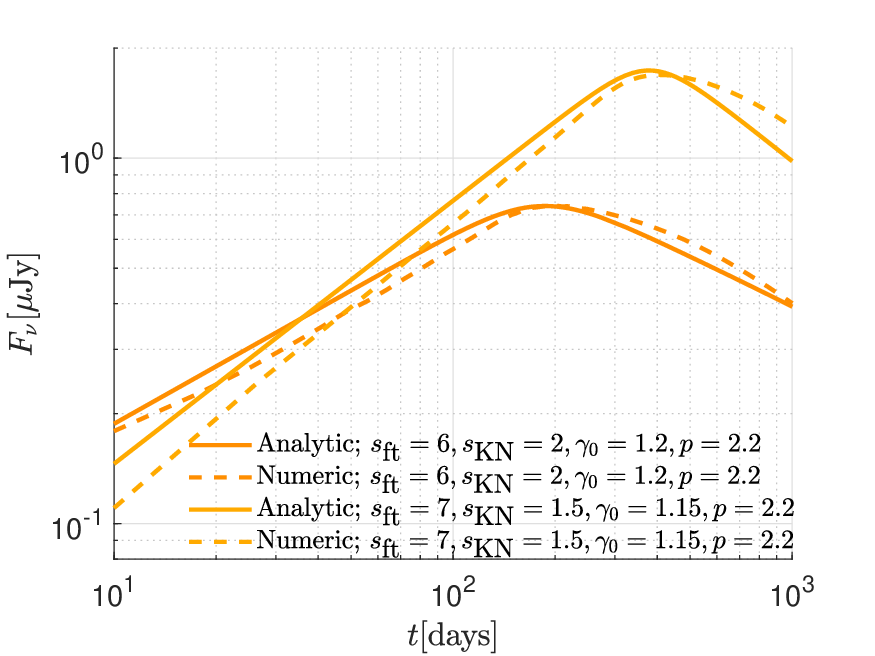}
\end{subfigure}
\begin{subfigure}[b]{0.5\textwidth}
    \centering
    \includegraphics[width=\columnwidth]{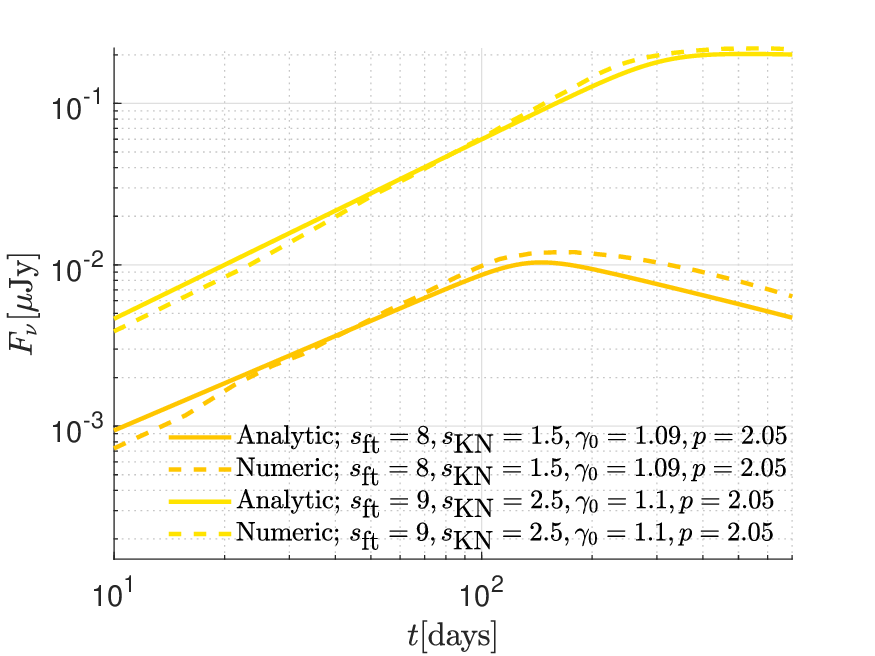}
    \end{subfigure}
    \caption{A comparison of the specific flux given analytically in \S~\ref{sec:results}, Eq.~(\ref{eq:betw}) (solid lines) and the specific flux obtained numerically (dashed lines), for different values of the dimensionless parameters $\{s_{\rm ft},s_{\rm KN},\gamma_0,p\}$ and $\varepsilon_e=10^{-1},\varepsilon_B=10^{-2},M_0=1.2\times10^{-5}M_\odot/5.2\times10^{-5}M_\odot/7.6\times10^{-7}M_\odot/1.1\times10^{-5}M_\odot$ (for $s_\text{ft}=6/7/8/9$) and $n=3\times10^{-2}\text{cm}^{-3}$. Results are shown for $\nu_{m},\nu_a<\nu=3\text{GHz}<\nu_{c}$ and observer distance of $D=100$Mpc.}
    \label{fig:compar_f}
\end{figure}

\begin{figure}
   \centering
    \includegraphics[width=\columnwidth]{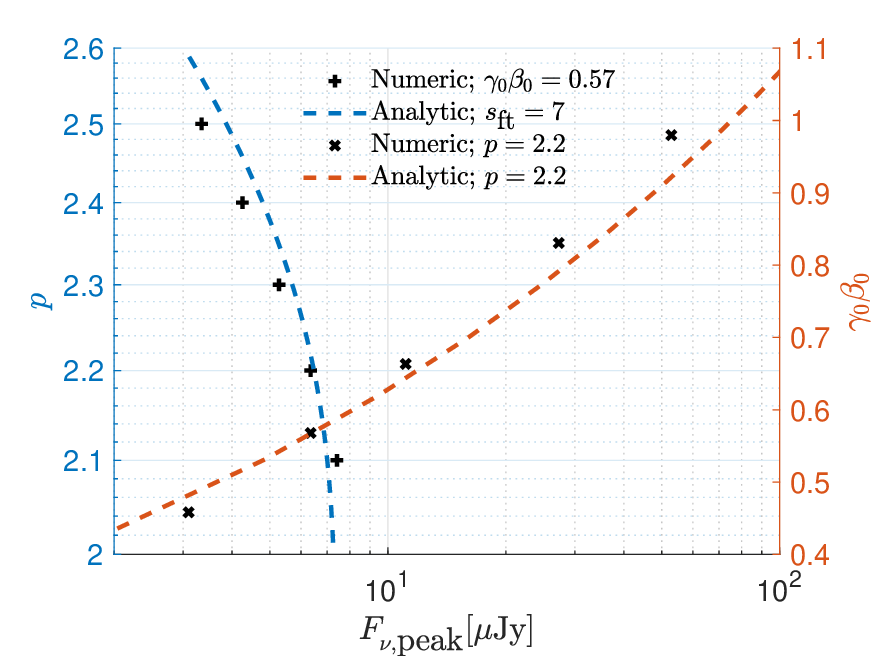}
    \caption{A comparison of the peak flux, $F_{\nu,\rm peak}$, given by the analytic model, Eq.~(\ref{eq:Fpeak}) (dashed lines), and that derived from numeric calculations ('+' and 'x' marks), as a function of $p$ and of $\gamma_0\beta_0$, for $M_0=10^{-4}M_\odot$ and $n=7\times10^{-2}\text{cm}^{-3}$. Results are shown for $\nu_{m},\nu_a<\nu=3\text{GHz}<\nu_{c}$ and observer distance of $D=100$Mpc.}
    \label{fig:compar_fp}
\end{figure}

\begin{figure}
   \centering
    \includegraphics[width=\columnwidth]{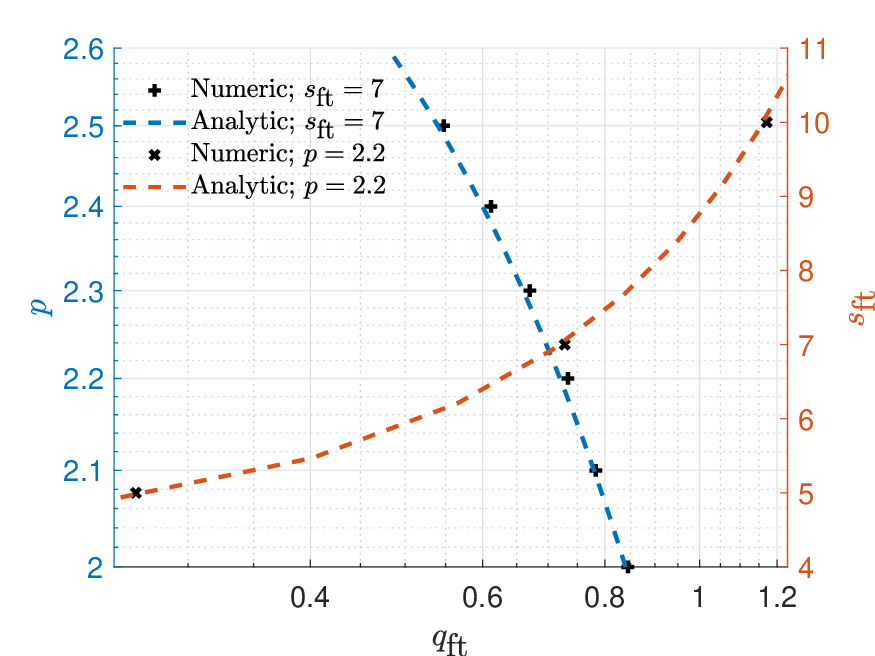}
    \caption{A comparison of the power-law index $q_{\rm ft}$ describing the $F_\nu\propto t^{q_{\rm ft}}$ time dependence of the specific flux before the peak, given by the analytic model, Eq.~(\ref{eq:an_end2}) (dashed lines), and that derived from numeric calculations ('+' and 'x' marks), as a function of $p$ and of $s_{\rm ft}$, for $M_R=10^{-6}M_\odot,n=3\times10^{-3}\text{cm}^{-3}$ and $\gamma_0=1.35$.}
    \label{fig:compar_q}
\end{figure}

\begin{figure}
    \centering
    \includegraphics[width=\columnwidth]{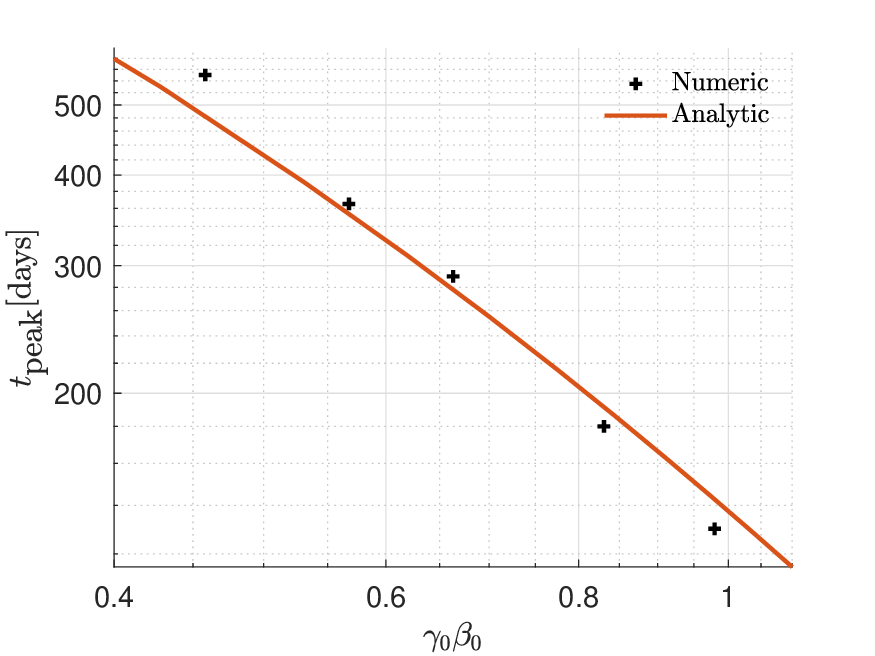}
    \caption{A comparison of the peak time, $t_\text{peak}$, given by the analytic model, Eq.~(\ref{eq:times}) (solid line), and that obtained by numeric calculations ('+' marks) for a range of  $\gamma_0\beta_0$ values, for $M_0=10^{-4}M_\odot$ and $n=7\times10^{-2}\text{cm}^{-3}$.}
    \label{fig:compar_tp}
\end{figure}

The emission of radiation was calculated, given a numeric solution for the flow fields, in a manner similar to that described at the opening of \S~\ref{sec:ntf}. As explained at the opening of \S~\ref{sec:model}, the dependence of the dynamics on the dimensional parameters $\{M_0, n, c\}$ follows directly from dimensional considerations, and the flow depends non-trivially only on the dimensionless parameters, $\{\beta_0,s_{\rm ft},s_{\rm KN}\}$. We have explored a wide range of these parameters $5<s_{\rm ft}<12,~1<s_{\rm KN}<3,~0.3<\beta_0<0.9$, and found $10$'s of percent agreement with our analytic approximation. We give below a few examples for the comparison of the full numeric calculation to the analytic approximation.  
\begin{enumerate}
\item In Fig.~\ref{fig:compar1} we compare the numeric solution for the forward shock velocity and radius with the results of the semi-analytic solution for $t<t_{\rm peak}$ described in \S~\ref{sec:dynamics}, and with the SvNT solution valid at late times. The good agreement supports the validity of both the numeric code and the semi-analytic solution of the dynamics.
\item In Fig.~\ref{fig:compar_f} we compare the specific flux obtained numerically with that given analytically in \S~\ref{sec:results}, Eq.~(\ref{eq:betw}). The agreement is within 
$10$'s of percent for a wide range of relevant values of $\{\beta_0,s_{\rm ft},s_{\rm KN}\}$.
\item In Fig.~\ref{fig:compar_fp} we compare $F_{\nu,\text{peak}}$ obtained numerically with the analytic expression of Eq. (\ref{eq:Fpeak}) for a range of $\gamma_0\beta_0$ and $s_{\rm ft}$ values. The numeric and analytic results are within 10's of percent agreement.
\item In Fig.~\ref{fig:compar_q} we compare the numeric logarithmic slope of the temporal dependence of the specific flux at $t<t_{\rm peak}$, $q_{\rm ft}=d\log F_\nu(t)/d\log t$, with the analytic expression of Eq.~(\ref{eq:an_end2}) for a range of $p$ and $s_{\rm ft}$ values, with a fixed $\beta_0$. The numeric and analytic results are within 10's of percent agreement.
\item In Fig.~\ref{fig:compar_tp} we compare $t_\text{peak}$ obtained numerically with the analytic expression of Eq.~(\ref{eq:times}) for a range of $\gamma_0\beta_0$ and $s_{\rm ft}$ values. The numeric and analytic results are within 10's of percent agreement.
\end{enumerate}

A comparison of the analytic results with numeric results for a power-law dependence of ejecta energy on momentum, see Eq.~(\ref{eq:profileE}), is given in \S~\ref{sec:earlier}. 


\section{Comparison to earlier work}
\label{sec:earlier}

We compare in this section our results with analytic results obtained in earlier papers. \cite{Nakar_2011} have considered a sub-relativistic shell of uniform $\beta$, and provided estimates of the peak time $t_{\rm peak}$, which is valid for $\beta\ll1$, and for the peak flux, which is based on extrapolation of expressions valid for highly relativistic ejecta. \cite{Kathirgamaraju_2019} (hereafter \citetalias{Kathirgamaraju_2019}) expanded the analysis of \cite{Nakar_2011}, considering a shell with a single power-law energy distribution, $E(>\gamma\beta)=E_R(\gamma\beta)^{-\alpha}$ for $\beta>\beta_0$. They have used the peak flux derived in \cite{Nakar_2011}, included an analysis of the dynamics to derive an expression for $t_{\rm peak}$ that takes into account relativistic effects and is valid for $\beta_0<0.5$, and used the results of \cite{Barniol_Duran_2015}, that were derived for highly relativistic ejecta, for the power-law index $q$ describing the $F_\nu\propto t^q$ time dependence of the specific flux at $t<t_{\rm peak}$.

Let us first note that the results of \citetalias{Kathirgamaraju_2019} are not self-consistent. At time $t<t_{\rm peak}$, when the reverse shock propagates through the $\beta>\beta_0$ ejecta, the emission should depend on the shell energy, $E_0=E(>\gamma_0\beta_0)$, and break/cutoff velocity, $\gamma_0\beta_0$, only through the combination $E_0(\gamma_0\beta_0)^\alpha=E_R$, since $E_R$ completely determines the dynamics at this stage- the shock propagates through the $E(>\gamma\beta)=E_R(\gamma\beta)^{-\alpha}=E_0(\gamma_0\beta_0/\gamma\beta)^\alpha$ ejecta and the existence of a break/cutoff at $\gamma_0\beta_0$ does not affect the flow up to $t=t_{\rm peak}$. See for example Eqs.~(\ref{eq:an_end1}) and~(\ref{eq:Fpeak})- the specific flux at $t<t_{\rm peak}$ depends only on $M_R$, corresponding to $E_R$, while the peak flux depends on $M_0$, corresponding to $E_0$. The expression that \citetalias{Kathirgamaraju_2019} provide for $F_\nu(t)$ depends on $E_0$ and $\gamma_0\beta_0$ separately, i.e. not only through the combination $E_0(\gamma_0\beta_0)^\alpha=E_R$. This implies that it would predict different fluxes at $t<t_{\rm peak}$ for ejecta with the same $E_R$, for which the predicted fluxes should be identical.

\begin{figure}
 \centering
\begin{subfigure}[b]{0.5\textwidth}
   \centering
    \includegraphics[width=\columnwidth]{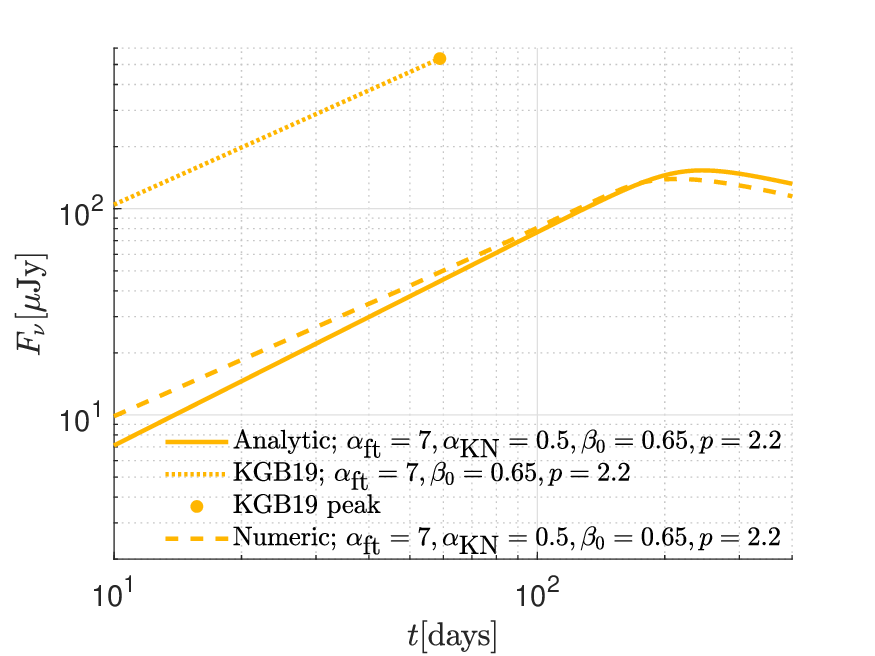}
\end{subfigure}
\begin{subfigure}[b]{0.5\textwidth}
    \centering
    \includegraphics[width=\columnwidth]{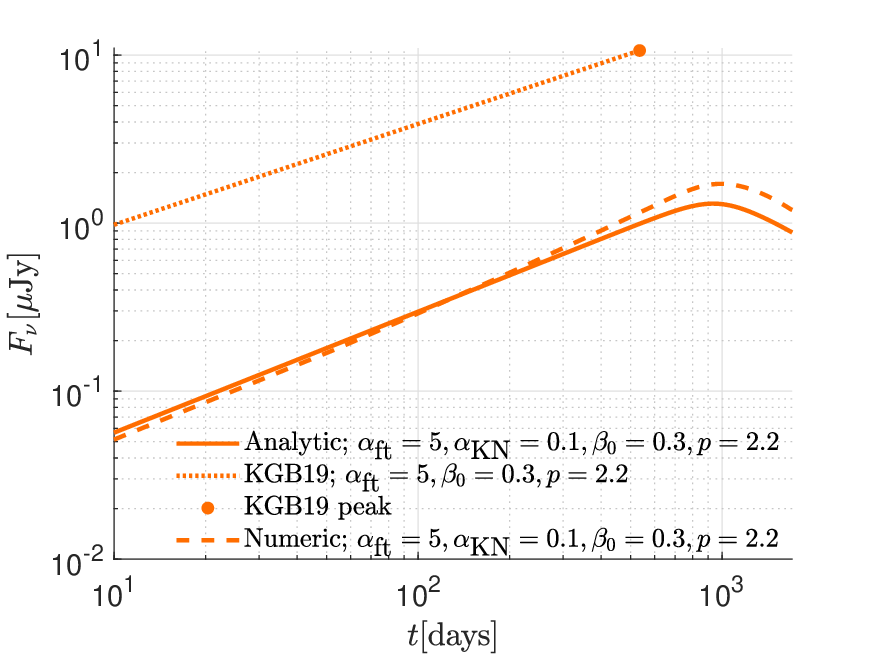}
    \end{subfigure}
    \caption{A comparison of the specific flux given analytically in \S~\ref{sec:results}, Eqs.~(\ref{eq:Eofv}) and (\ref{eq:betw}) (solid lines) with the specific flux obtained numerically (dashed lines) and that given analytically in \citetalias{Kathirgamaraju_2019}, for ejecta profiles given by Eq.~(\ref{eq:profileE}) with different values of the dimensionless parameters $\{\alpha_{\rm KN},\alpha_{\rm ft},\beta_0,p\}$ and $\varepsilon_e=10^{-1},\varepsilon_B=10^{-2},E_0=10^{50}\text{erg}/10^{49}\text{erg}$ (upper/lower panel) and $n=3\times10^{-2}\text{cm}^{-3}$. Results are shown for $\nu_{m}/\nu_a<\nu=3\text{GHz}<\nu_{c}$ and observer distance of $D=100$Mpc.}
    \label{fig:compar_en}
\end{figure}

In Fig.~\ref{fig:compar_en} we compare the specific flux given by our analytic model, using Eq. (\ref{eq:Eofv}), for an ejecta with a broken power-law dependence of ejecta energy on momentum, Eq.~(\ref{eq:profileE}), with the specific flux obtained by our numeric calculations and that given by \citetalias{Kathirgamaraju_2019}. Our analytic expressions provide a good approximation (within 10's of percent) for the numerical results for different values of $\alpha_{\rm ft}$ ($5,7$) and $\beta_0$ ($0.3,0.65$). The analytic expression of \citetalias{Kathirgamaraju_2019}, based on \cite{Nakar_2011}, overestimates the specific flux by typically an order of magnitude, and significantly underestimate the peak time for $\beta=0.65$.
The ratio of the peak flux given by Eqs.~(\ref{eq:Fpeak}-\ref{eq:gfactor}) and~(\ref{eq:Eofv}) and that given by \citetalias{Kathirgamaraju_2019} is (pre-factor given for $p=2.2$)
\begin{equation}\label{eq:fpeak-ratio}
  \frac{F_{\nu,\text{peak}}^{\text{KGB19}}}{F_{\nu,\text{peak}}}\approx 6 \gamma_0^{2.5-2.2p}\beta_0^{0.3p-1}.
\end{equation}

\begin{figure}
    \centering
    \includegraphics[width=\columnwidth]{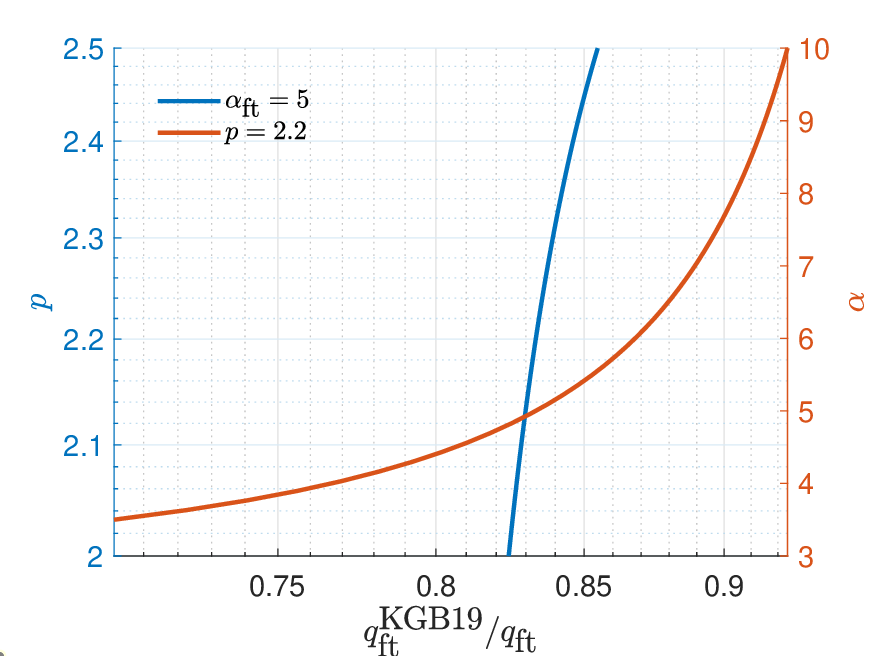}
    \caption{The ratio between the logarithmic slope $q_{\rm ft}=d\log F_\nu(t)/d\log t$ of the temporal dependence of the specific flux at $t<t_{\rm peak}$ obtained by \citetalias{Kathirgamaraju_2019} and our result, given by Eqs.~(\ref{eq:Eofv}) and~(\ref{eq:an_end2}), for different values of $\alpha_{\rm ft}$ and $p$.}
    \label{fig:compar_en_q}
\end{figure}

\begin{figure}
    \centering
    \includegraphics[width=\columnwidth]{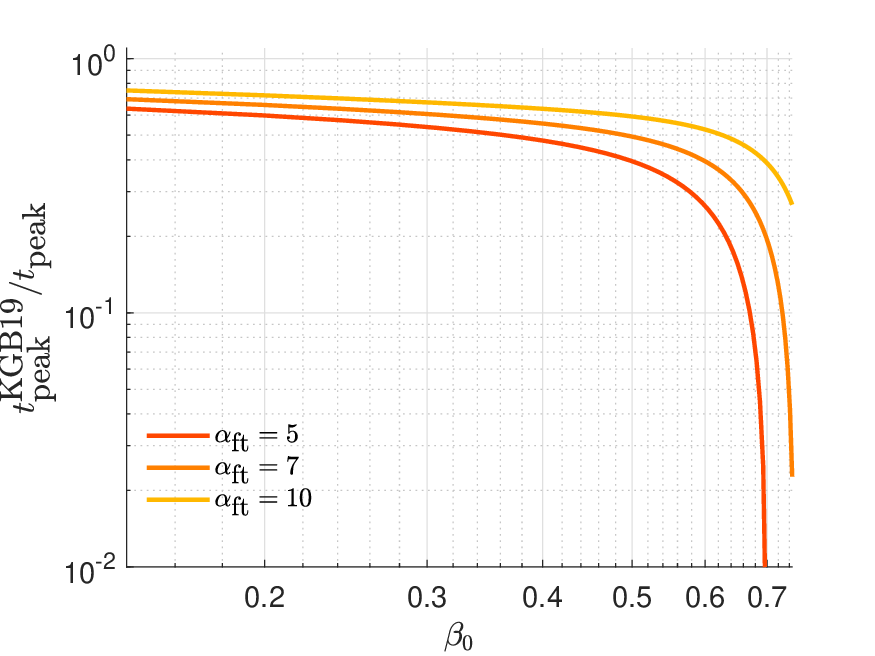}
    \caption{The ratio between the peak time, $t_{\rm peak}$, obtained by \citetalias{Kathirgamaraju_2019}, and our result, given by Eqs.~(\ref{eq:Eofv}) and~(\ref{eq:times}).}
    \label{fig:compar_en_tp}
\end{figure}

Fig.~\ref{fig:compar_en_q} shows the ratio between the logarithmic slope $q_{\rm ft}=d\log F_\nu(t)/d\log t$ of the temporal dependence of the specific flux at $t<t_{\rm peak}$ obtained by \citetalias{Kathirgamaraju_2019}, $q_{\text{ft}}^{\text{KGB19}}=(3\alpha-6(p-1))/(8+\alpha_{\rm ft})$, and our result, given by Eqs.~(\ref{eq:an_end2}) and~(\ref{eq:Eofv}), for different values of $\alpha_{\rm ft}$ and $p$. Fig.~\ref{fig:compar_en_tp} shows the ratio between the peak time, $t_{\rm peak}$, obtained by \citetalias{Kathirgamaraju_2019}, $t_{\text{peak}}^{\text{KGB19}}=3.3\left(\frac{E_{0,51}}{n_{-2}}\right)^{\frac{1}{3}}\beta_0^{-\frac{2}{3}} \left(\frac{2+\alpha_{\rm ft}}{\beta_{0}(5+\alpha_{\rm ft})}-1 \right)\text{years}$, and our result, given by Eqs.~(\ref{eq:times}) and~(\ref{eq:Eofv}). As expected, the \citetalias{Kathirgamaraju_2019} expression provides a good approximation for $t_{\rm peak}$ for $\beta_0<0.5$.


\section{GW170817 KN afterglow}
\label{sec:GW170817}

\begin{figure}
    \centering
    \includegraphics[width=\columnwidth]{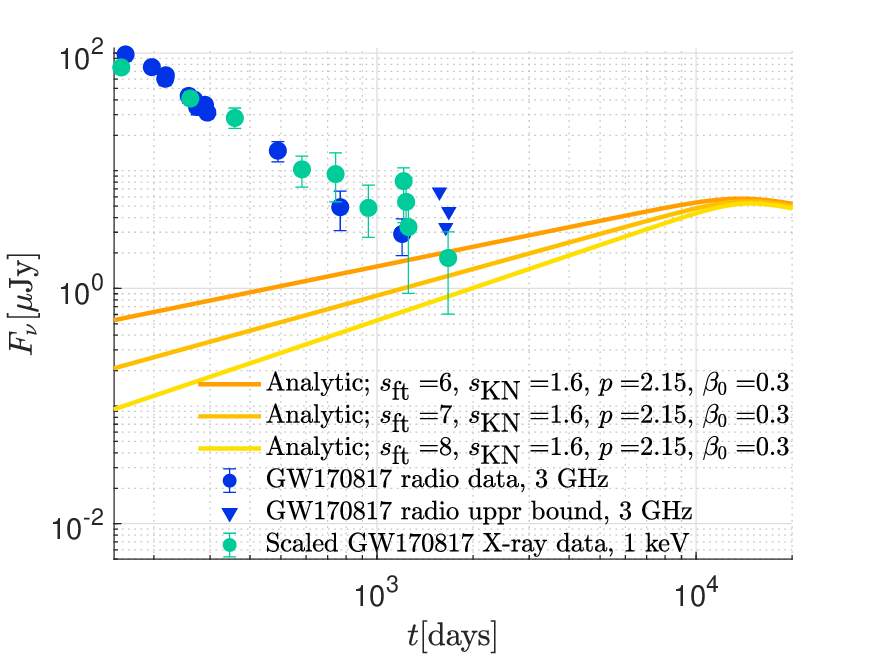}
    \caption{The radio and X-ray observations (symbols) of the electromagnetic counterpart of GW170817 \citep{Troja_2021,Troja_2022,Oconnor_2022,Balasubramanian_2021,Balasubramanian_2022},
    and the radio flux (solid lines) predicted to be produced by a "fast-tail" extension, $M(>\gamma\beta)\propto(\gamma\beta)^{-s_{\rm ft}}$ at $\beta>\beta_0=0.3$, of the mass distribution inferred from observations at lower velocity, $M(>\gamma\beta)\propto(\gamma\beta)^{-1.6}$ at $0.1<\gamma\beta<0.3$ and a total mass of $0.05M_\odot$ \citep{Waxman_2018}. X-ray data are scaled by a factor $3\times10^8$ (corresponding to electron spectral index $p=2.15$). Model results are shown for $\varepsilon_e=10^{-1},\varepsilon_{B}=0.5\times10^{-2}$ and $n=10^{-3}{\rm cm^{-3}}$.}
    \label{fig:GW170817_data}
\end{figure}

The radio and X-ray observations of the electromagnetic counterpart of GW170817 are shown in figure~\ref{fig:GW170817_data}. The measurements show a growing tension with the predictions of expanding jet models, but clear evidence for re-brightening associated with a fast tail is not (yet) present \citep{Hajela_2022,nedora_2021a,Balasubramanian_2021,Balasubramanian_2022,Troja_2022,Oconnor_2022}.

As noted in the introduction, the early evolution of the photospheric radius is consistent with a power-law mass distribution, $M(>\gamma\beta)\propto(\gamma\beta)^{-s_{\rm KN}}$ with $s_{\rm KN}\approx1.6$ at $0.1<\gamma\beta<0.3$ and a total mass of $\approx 0.05M_\odot$ \citep{Waxman_2018}. We consider the non-thermal emission that would be produced by a "fast-tail" extension of this power low to higher velocities, with $s_{\rm ft}$ in the range of 6-8 for $\beta>\beta_0=0.3$ ($M_0=8\times10^{-3}M_\odot$). Figure~\ref{fig:GW170817_data} shows the non-thermal specific flux predicted by our model for typical values expected for collisionless shocks, $\varepsilon_e=10^{-1},\varepsilon_B=0.5\times10^{-2},p=2.15$, and conservatively low density, $n=10^{-3}{\rm cm^{-3}}$. The existence of a fast tail is expected to produce detectable radio and X-ray fluxes over a time scale of $\sim10^4$days.

\section{Summary}
\label{sec:summary}
We have derived analytic expressions for the non-thermal synchrotron emission from spherical ejecta with broken power-law mass or energy distributions given by Eqs.~(\ref{eq:profile}) or~(\ref{eq:profileE}). The analytic model equations are given in \S~\ref{sec:results}. The analytic model results were compared in \S~\ref{sec:numeric} and in \S~\ref{sec:earlier} to the results of 1D numeric calculations for a wide range of ejecta parameter values characteristic of merger calculation results- a "shallow" mass (energy) distribution, $1<s_{\rm KN}<3$, for the bulk of the mass (at $\gamma\beta\approx 0.2$), and a steep, $s_{\rm ft}>5$, "fast tail" mass (or energy) distribution. The analytic model provides an accurate (to 10's of percent) description of the evolution of the specific flux obtained in the numeric calculations, including at the phase of deceleration to sub-relativistic expansion- see figures~\ref{fig:compar_f}-\ref{fig:compar_en}. This is a significant improvement over earlier results, based on extrapolations of results valid for $\gamma\beta\gg1$~ and $\ll1$ to $\gamma\beta\approx1$, which overestimate the specific flux by an order of magnitude for typical parameter values and do not describe the emission beyond the time of peak flux (see discussion in \S~\ref{sec:earlier} and figures~\ref{fig:compar_en}-\ref{fig:compar_en_tp}). 

The key results are the time and magnitude of the peak flux (at $\nu<\nu_c$), given by
\begin{equation}
    \begin{aligned}
         t_\text{peak}&= 550g(\beta_0)\left(\frac{M_{0,-4}}{n_{-2}}\right)^{\frac{1}{3}}\text{days},\\
       F_{\nu,\rm peak}&=10D_{26.5}^{-2}\varepsilon_{e,-1}^{p-1}f_{\rm ft}(p)
    \varepsilon_{B,-2}^{\frac{p+1}{4}}n_{-2}^{\frac{p+1}{4}}M_{0,-4}\nu_{9.5}^\frac{1-p}{2}g_q \mu\text{Jy},\\
    q_{\rm ft}&=\frac{4.5-7.5p+3s_{\rm ft}}{5.5+s_{\rm ft}},\quad q_{\rm KN}=\frac{7.5-7.5p+3s_{\rm KN}}{4.7+s_{\rm KN}},
        \end{aligned}
\end{equation}
where $q_{\rm ft/\rm KN}$ is the index describing the $F_\nu\propto t^{q_{\rm ft/\rm KN}}$ time dependence of the specific flux before/after the peak (for $6<s_{\rm ft}<8$, $1<s_{\rm KN}<2.5$ and $2<p<2.2$ we find $0.5<q_{\rm ft}<1$ and $-1<q_{\rm KN}<0$), and
\begin{equation}
    \begin{aligned}
    g(\beta_0)&=\frac{1.5-\sqrt{0.25+2\beta_0^2}}{\gamma_0^{\frac{1}{3}}\beta_{0}},\\
        f_{\rm ft}(p)&=960\times8.8^{-p}(2.5-0.7p)(p-1)^{2-p}(l_p)^{p-1}\approx1,\\
    g_q&=2.3^{q_{\rm ft}} (\gamma_0\beta_0)^{s_{\rm ft}\left(1-\frac{q_{\rm ft}}{3}\right)} (g(\beta_0))^{q_{\rm ft}}\approx1.2(\gamma_0\beta_0)^{2.2p-0.5}.
    \end{aligned}
\end{equation}
Here, $\gamma_0\beta_0$ is the break between the fast part of the ejecta ($s_\text{ft}>5$) and the slower part of the ejecta ($1<s_\text{KN}<3$), see Eq.~(\ref{eq:profile}), $M_0=10^{-4}M_{0,-4}M_\odot$ is the (total) mass of the fast tail, $n=10^{-2}n_{-2}{\rm cm}^{-3}$ is the ISM density, $\varepsilon_{e}=10^{-1}\varepsilon_{e,-1}$ and $\varepsilon_{B}=10^{-2}\varepsilon_{B,-2}$ are the fractions of internal energy carried by electrons and magnetic fields, and $p$ is the electrons power-law index, $dn_e/d\gamma_e\propto \gamma_e^{-p}$. A routine calculating the specific flux based on the analytic results presented in this paper is available at \url{https://gitlab.com/Gilad_Sadeh/sadeh-et-al-2022}. 

The flux peaks when the reverse shock crosses the steep fast tail, $\gamma\beta>\gamma_0\beta_0$, part of the ejecta into the shallower part at lower velocity. It may be detected in the radio and X-ray bands, e.g. by the Very Large Array radio telescope (with a limiting sensitivity of $\sim 5\mu$Jy for $1-10$GHz frequencies) and by the Chandra X-ray observatory (with a limiting sensitivity of $\sim 10^{-4}\mu$Jy for $0.3-10$keV photons for $\sim10$ks exposure time).

Our analytic model will enable a reliable direct inference of the values and uncertainties of model parameters from observations. While numeric calculations may provide higher accuracy predictions, the analytic model has a great advantage when inferring model parameters from observations, since it does not require detailed resource consuming calculations for each point in the model's multi-parameter space. 

The late time non-thermal emission of GW170817 was considered in $\S$\ref{sec:GW170817}. Figure~\ref{fig:GW170817_data} shows the radio and X-ray flux that would be produced by a fast tail, extending the observationally inferred $M(>\gamma\beta)\propto (\gamma\beta)^{-s_{\rm KN}}$ with $s_{\rm KN}\approx 1.6$ for $\beta<0.3$ \citep{Waxman_2018}, to higher velocity as $M(>\gamma\beta)\propto (\gamma\beta)^{-s_{\rm ft}}$ with a range of $s_{\rm ft}$ values. Using typical values for $\varepsilon_{e,B}$ and $p$, and adopting a conservatively low density, $n=10^{-3}{\rm cm^{-3}}$, we find that the existence of a fast tail is expected to produce a detectable (few $\mu$Jy) radio and ($10^{-15}{\rm erg/cm^2s}$) X-ray fluxes over a time scale of $\sim10^4$days.

\section*{Acknowledgements}
We thank Jonathan Morag and Arnab Sarkar for insightful discussions during the conduction
of this work.
Eli Waxman's research is partially supported by ISF, GIF and IMOS grants.

\section*{Data Availability}
The data underlying this article will be shared following a reasonable request to the corresponding author. A routine calculating the specific flux based on the analytic results presented in this paper is available at \url{https://gitlab.com/Gilad_Sadeh/sadeh-et-al-2022}.


\bibliographystyle{mnras}
\bibliography{bib} 




\appendix

\section{Power-law approximations}
\label{appendix_power-law}

The $(\gamma\beta)^\lambda$ power-law approximations used for the derivation of the non-thermal specific flux produced by the fast tail in Eqs. (\ref{eq:an_end2}-\ref{eq:an_end3}) are (accurate to 10\% for $0.3<\gamma\beta<2.5$)
\begin{equation}
    \nu_c\propto \frac{\left(1.5-\sqrt{2\beta^2+0.25}\right)}{\gamma^{\frac{1}{2}}(\gamma-1)^{\frac{3}{2}}}\approx (\gamma\beta)^{-3.9},
\end{equation}

\begin{equation}
L_\nu (\nu>\nu_c)\propto\frac{(\gamma-1)^{(p+2)/2}\gamma^{\frac{3-p}{2}}\beta^{3}}{\left(1.5-\sqrt{2\beta^2+0.25}\right)^{\frac{p+4}{2}}}\approx 7.7(1.25-0.35p)(\gamma\beta)^{2.5p+2},
\end{equation}
\begin{equation}
L_\nu(\nu<\nu_c)\propto\frac{(\gamma-1)^{(p+3.5)/2}\gamma^{\frac{3.5-p}{2}}\beta^{3}}{\left(1.5-\sqrt{2\beta^2+0.25}\right)^{\frac{p+5}{2}}}\approx 7.7(1.25-0.35p)(\gamma\beta)^{2.5p+4}.
\end{equation}
The $(\gamma\beta)^\lambda$ power-law approximations used for the derivation of the non-thermal specific flux produced by the slower part of the ejecta in Eqs. (\ref{eq:an_end1_2}-\ref{eq:an_end3_2}) are (accurate to 10\% for $0.1<\gamma\beta<0.7$)
\begin{equation}
    \nu_c\propto \frac{\left(1.5-\sqrt{2\beta^2+0.25}\right)}{\gamma^{\frac{1}{2}}(\gamma-1)^{\frac{3}{2}}}\propto (\gamma\beta)^{-3.3},
\end{equation}
\begin{equation}
L_\nu(\nu>\nu_c)\propto\frac{(\gamma-1)^{(p+2)/2}\gamma^{\frac{3-p}{2}}\beta^{3}}{\left(1.5-\sqrt{2\beta^2+0.25}\right)^{\frac{p+4}{2}}}\propto(\gamma\beta)^{2.5p+0.6},
\end{equation}
\begin{equation}
L_\nu(\nu<\nu_c)\propto\frac{(\gamma-1)^{(p+3.5)/2}\gamma^{\frac{3.5-p}{2}}\beta^{3}}{\left(1.5-\sqrt{2\beta^2+0.25}\right)^{\frac{p+5}{2}}}\propto(\gamma\beta)^{2.5p+2.2}.
\end{equation}

\bsp	
\label{lastpage}
\end{document}